\documentclass[a4paper,11pt]{article}
\usepackage{pos}
\usepackage{graphicx}   
\usepackage{amsmath}    
\usepackage{hyperref}
\usepackage{soul}

\newcommand{\oversim}[2]{\protect{\mbox{\lower0.5ex\vbox{%
  \baselineskip=0pt\lineskip=0.2ex
  \ialign{$\mathsurround=0pt #1\hfil##\hfil$\crcr#2\crcr\sim\crcr}}}}}
\newcommand{\simgreat}{\mbox{$\,\mathrel{\mathpalette\oversim>}\,$}} 
\newcommand{\simless} {\mbox{$\,\mathrel{\mathpalette\oversim<}\,$}} 

\title{The many tensions with dark-matter based models and implications on the   nature of the Universe}
\ShortTitle{Tensions with dark-matter models}

\author*[a,b]{Pavel Kroupa}
\affiliation[a]{
Helmholtz-Institut f\"ur Strahlen- und Kernphysik,
Universit\"at Bonn, Nussallee 14-16, 53115 Bonn, Germany}
\affiliation[b]{
  Astronomical Institute,
  Charles University, V Holesovickach 2, 18000 Praha, Czech Republic}

\emailAdd{pkroupa@uni-bonn.de}
\author[c,d]{Eda Gjergo}
\affiliation[c]{School of Astronomy and Space Science, Nanjing
  University, Nanjing 210093, People’s Republic of China}
\affiliation[d]{Key Laboratory of Modern Astronomy and Astrophysics (Nanjing University),
  Ministry of Education, Nanjing 210093, People’s Republic of China}

\author[a]{Elena Asencio}
\author[a]{Moritz Haslbauer}
\author[a]{Jan Pflamm-Altenburg}
\author[a]{Nils Wittenburg}
\author[b]{Nick Samaras}
\author[a]{Ingo Thies}
\author[a]{Wolfgang Oehm}

\abstract{ The current standard models of cosmology (SMoC) --
  specifically $\Lambda$CDM and warm dark matter models -- served for
  a few decades as the basis of research in astronomy and cosmology,
  and have been studied extensively.  However, fundamental tensions
  between observations and theory have emerged. This updated review
  has two purposes: to explore new tensions that have arisen in recent
  years, compounding the unresolved tensions from previous studies,
  and to use the shortcomings of the current theory to guide the
  development of a successful model.

  In any representative volume, more than 90~per cent of all galaxies
  have thin, extended star-forming ancient disks.  But these
  structures are too fragile to withstand the repeated mergers that
  dark matter would induce.  According to SMoC cosmological
  simulations, galaxies in the Local Group around the Mpc scale should
  be distributed in a spheroidal configuration. Observations show that
  they are instead arranged in thin planes. This poses major questions
  on the nature and dynamical history of the Local Group.
  Furthermore, there exist mutually correlated planes of satellite
  dwarf galaxies located around the Andromeda and Milky Way
  galaxies. These planes may have been created by the tidal forces
  generated by a previous encounter between the two galaxies. Also the
  configuration of the nearby M81~group poses challenges to the
  SMoC. None of these structures could exist in the presence of dark
  matter, because dynamical dissipation would cause the galaxies to
  merge within a Gyr time scale. In addition, the El Gordo galaxy
  cluster has been observed at a redshift at emission of
  $z_{\rm e}=0.87$ to already have reached a mass of
  $\approx 2\times 10^{15}\,M_\odot$, being impossible in the SMoC.
  In the light of the growing evidence for large-($>$few
  hundred~Mpc)-scale inhomogeneities, we have shown that the Hubble
  Tension is simply caused by the matter bulk flows of the large scale
  structure.  These observations suggest that the Universe is more
  inhomogeneous and that structures grow more rapidly than what is
  allowed by the SMoC on the Gpc~scale. Novel tensions have emerged
  from observations of the early Universe. For instance, galaxies of
  mass $10^9-10^{10}\,M_\odot$ have been detected in the redshift
  range of $10\simless z_{\rm e} \simless 20$, indicating a faster
  galaxy formation than predicted in the SMoC. An independent
  indication for such early galaxy formation comes from the downsizing
  timescales of early-type galaxies and their associated rapid
  formation of central super-massive black holes (SMBHs).

  We discuss a few candidate models of cosmology from the literature,
  but most fail on all or a number of the above problems. Given the
  nature of the tensions, the real Universe needs to be described by a
  model in which gravitation is effectively stronger than
  Einsteinian/Newtonian gravitation at accelerations below Milgrom's
  acceleration scale. Interestingly, Milgromian dynamics (MOND)
  coincides with the generalized Poisson equation given by the
  non-linear p-Laplacian when $p=3$ with $p=2$ providing the standard
  Newtonian Poisson equation. A promising model that solves several of
  the above tensions is $\nu$HDM. While it embraces MOND, eliminating
  the need for dark matter, it retains dark energy and consequently
  the SMoC expansion history.  However galaxy formation appears to
  occur too late in this model, model galaxy clusters reach too large
  masses, and the mass function of model galaxy clusters is too flat
  and thus top-heavy in comparison to the observed mass function.
    
  The above mentioned evidence casts doubts on the viability of dark
  matter and dark energy as fundamental components of the Universe,
  with severe consequences. Specifically, the Hot Big Bang Theory
  cannot provide a good fit to the CMB power spectrum without invoking
  both of these components. Consequently, inflation -- introduced to
  justify why causally-disconnected regions of the CMB should be
  homogeneous on a flat geometry -- would also cease to be needed. The
  models that have been simulated to-date with these boundary
  conditions appear to not be able to generate structure rapidly
  enough to be consistent with the high-redshift JWST observations.
  Given all the noted anomalies, the classes of models that relax
  these boundary conditions should be explored.  }

\FullConference{%
  Corfu Summer Institute 2022 ``Workshop on Tensions in Cosmology'',\\
  7 -- 12 September, 2022\\
  Corfu, Greece}

 \tableofcontents

\begin{document}
\maketitle

\section{Introduction} \label{sec:intro}

The development of the $\Lambda$CDM and $\Lambda$WDM models of
cosmology, based on the required existence of cold or warm dark matter
particles (CDM and WDM, respectively), rightly counts as one of the
greatest achievements of physics (e.g. \cite{Peebles1993, Peacock1999,
  Schneider2015, Turner2022}).  Hereinafter, to simplify terminology,
we will refer to $\Lambda$CDM and $\Lambda$WDM (e.g. \cite{Rose+2023})
jointly as the currently favoured standard model of cosmology, or SMoC
(this includes fuzzy dark matter, see Sec.~\ref{sec:motivation}).

The SMoC accounts for the observed expansion of the Universe, the
cosmic microwave background (CMB) and allows the numerical
quantification of the formation of model galaxies, model galaxy
clusters as well as the filamentary distribution of matter on Mpc and
Gpc scales.

Recent high-resolution simulations use a range of algorithms to
compute gravitational potentials, most commonly opting for
smooth-particle hydrodynamics or adaptive mesh refinement
methods. These simulations also incorporate additional physical
processes beyond gravity, including hydrodynamics, star formation, and
feedback from both stars and massive black holes. As a result, it is
now possible to directly compare simulated galaxy populations with the
respective observations. Yet, a growing body of observational
constraints -- covering scales from open star clusters to the
cosmological horizon -- have highlighted several discrepancies. Some
of these discrepancies are especially troubling.  Given these, it is
useful to develop and study different cosmological models in order to
improve our appreciation of the current SMoC, while perhaps also
finding directions that might lead to fruitful advances away from
it. In this process, it is important to develop models that allow the
simulation of structures down to galaxies. This requires, like the
SMoC, the alternative model to provide equations of motion that can be
integrated in a computer to solve for structure formation on all
scales.  The emphasis here is thus set on computable models of
structure formation down to the scales of galaxies or smaller, rather
than a more theoretical discussion of abstract models.  We emphasise
that any theoretical approach away from the current SMoC needs to be
guided by the constraints noted here, namely by the non-existence of
dark matter particles in galaxies and that the real Universe appears
to be structured on all scales that have been probed so far and up to
the horizon scale, and needs to be able to have formed galaxies with
baryonic masses $\simgreat10^9\,M_\odot$ by a redshift
$10 \simless z_{\rm e}\simless 20$ subject to these observations being
spectroscopically confirmed (see possible tension pT3 below).

Two such new model universes that can be simulated in a computer are
presently being studied in detail. The first is the Angus model of
cosmology (AMoC). It rests on a traditional or conservative approach
by adopting the same expansion history as the SMoC. The AMoC is based
on the Milgromian equations of motion, rather than the Newtonian
ones. A less-conventional or neoconservative
model\footnote{``Neoconservative'' in terms of relying on less-exotic
  physics than the current SMoC, exotic physics encompassing
  inflation, dark matter particles and dark energy.}, the Bohemian
model of cosmology (BMoC), is now being studied in Bonn, Prague
and Nanjing and will be reported elsewhere.

This contribution can be viewed as an update on the previous
assessments of the SMoC (\cite{Kroupa+2010, Kroupa2012,
  Kroupa2015}), outlining the motivation for studying non-dark-matter
based models (Sec.~\ref{sec:motivation}).  Some such alternatives are
discussed in Sec.~\ref{sec:alternatives}.  A particular case, the
AMoC, allows the computation of galaxies and larger structures and is
briefly described in Sec.~\ref{sec:AMoC}.  The conclusions are
available in Sec.~\ref{sec:conclusions}. 
Appendix~\ref{sec:MOND} provides a short introduction to MOND.

\section{Motivation}
\label{sec:motivation}

\subsection{Standard (dark-matter-based) Models of Cosmology}
\label{sec:SMoC}

In the following the postulates and hypotheses underlying the SMoC are
discussed with the aim of confronting these with those underlying the
AMoC (Sec.~\ref{sec:AMoC}) and for paving the way for constructing the
BMoC.

The current SMoC rests on five fundamental postulates: {\bf (A1)} The
locally observed laws of physics (as embodied in the standard model of
particle physics, the SMoPP) are valid everywhere.  {\bf (A2)}
Cosmological redshifts are given by expanding space along the line of
sight.\footnote{This is a fundamental postulate not further mentioned
  and underlies all models discussed here. A cosmological redshift
  does not signify a velocity but the photon wavelength is stretched
  through the space expansion. A galaxy at a high redshift is
  stationary apart for its peculiar velocity, but it appears to be
  receding due to the expansion of space along the line of sight.}
{\bf (A3)} The Equivalence Principle applies and Einsteinian
gravitation is universally valid. {\bf (A4)} The Cosmological
Principle applies, i.e., the global coordinates of the Universe are
characterized by homogeneity and isotropy. Thus, the large scale
structure of the Universe is represented by the
Friedmann-Lemaitre-Robertson-Walker (FLRW) metric ,
$ds^2=c^2\,dt^2 - a_{\rm e}(t)^2\left(dx^2+dy^2+dz^2\right)$, where
$c$ is the speed of light, $dx, dy, dz$ are co-moving
three-dimensional distance intervals and
$a_{\rm e}(t)= \left(1+z_{\rm e}\right)^{-1}$ is the time-dependent
scale factor with $z_{\rm e}(t)$ being the redshift
(e.g. \cite{Schneider2015}), so that a present-day separation of 1~Mpc
spanned $48\,$kpc at redshift $z_{\rm e}=20$.  {\bf (A5)} All baryons
and leptons have a primordial origin, i.e. they formed in the
electroweak epoch ($10^{-36}$~s to $10^{-12}$~s) before the four
fundamental forces separated \citep{Gibbons+1983, RiottoTrodden1999}.

Given these five postulates, auxiliary hypotheses need to be invoked to
account for some observations: {

  \bf (AH1)}~A {\it cosmic inflationary epoch} during the earliest
stages of the Big Bang (e.g. \cite{Linde2001}) is needed to ensure for
the homogeneity and isotropy of the observed CMB.  Inflation was
introduced \citep{Guth1981} to explain why causally-disconnected
regions of the cosmic microwave background (CMB) should be
homogeneous, and why the geometry of the Universe is flat.  Given that
the observed Universe appears to adhere to exactly the same
SMoPP\footnote{Variations of the fundamental constants are being
  searched for, see \cite{PeriSkara2022} for a recent review. See also
  \cite{Liu+2023} for a suggestion to measure the possible variation
  of $c$.}, inflation also ensures causal connection of every part of
the Universe. The model universe is implied to be flat and this is
supported by the angular distance of the three baryonic oscillation
peaks in the observed CMB that serve as a standard ruler
\citep{Planck2018}.\footnote{Possible evidence for a small curvature
  has emerged \citep{DiValentino+2021a}} {\bf (AH2)}~{\it Dark matter
  particles} need to make up about 95~per cent of all of the
matter. This is motivated by the gravitational anomalies observed in
galaxies that appear as missing mass in the adopted gravitational
theory, as well as through the three major acoustic peaks observed in
the CMB power spectrum. It is important to note that there is no
motivation from the SMoPP for the existence of additional such
particles. In order to not interfere with the tight experimental
constraints available for the SMoPP, the dark matter needs to be made
of particles that are not in the SMoPP and that only interact with
SMoPP particles via gravitation and at most through the weak
force. Dark matter particles can be cold, warm or fuzzy, depending on
their kinetic energy and Compton wavelength set during the matter
creation era during inflation. Ultimately, each version needs to
account for the observationally deduced dark matter content of dwarf
(diameter $<500\,$pc) galaxies constraining the Compton wavelength to
be small such that the dynamics of galaxies is not much affected in
comparison to standard CDM particles with a negligible Compton
wavelength \citep{RogersPeiris2021, DalalKravtsov2022,
  Khelashvili+2022}. {\bf (AH3)}~With A1--A5 and AH1 \& AH2 an
expanding model universe is obtained that slows with time through its
eigengravity. Observations lead to the result that the real Universe
expands more rapidly than thusly expected \citep{YoshiiPeterson1995,
  Perlmutter+1998, Schmidt+1998, Riess+1998}. This discrepancy can
only be remedied by invoking, as a third auxiliary hypothesis, that
{\it dark energy} (DE) accounts for about 67~per cent of the total
energy content of the Universe.  Adding such a constant term is
allowed by Einstein's theory of general relativity.  DE, an energy
density, leads to a time-increasing rate of expansion, starting about
$5\,$Gyr ago.  The expansion history of the SMoC is conveniently given
by eq.~36 in \cite{BanikZhao2016}.  Postulates A1-A5 together with AH2
imply that Newtonian gravitation can be used to model the evolution of
structures in an isotropic and homogeneous background expansion
described by the Friedmann equations for times later than
$z_{\rm e}=1100$.

Given the above, six parameters define the current (today, at time
$t=t_0$) state of the model universe in terms of its expansion rate,
mass components, initial density fluctuations and curvature (see
\cite{Planck2018}): The Hubble-Lemaitre constant
$H_0\approx 67.4\pm 0.5\,{\rm km}\,{\rm s}^{-1}\,{\rm Mpc}^{-1}$, the
baryon density parameter $\Omega_{\rm b,0} \approx 0.05$, the dark
matter density parameter $\Omega_{\rm c,0} \approx 0.26$ (leading to
the matter density parameter $\Omega_{\rm m,0}\approx 0.31$) and
the dark energy density parameter $\Omega_{\Lambda,0} \approx 0.69$
which follows from the condition $\Omega_{\rm tot}=1$ for a flat
universe. The age of this model universe is $t_0 = 13.8\,$Gyr. 
Note that as the SMoC expands the fraction of its energy
content that is in dark energy increases, while the total energy
content in matter is fixed.  The scalar spectral index,
$n_{\rm s}\approx 0.97$, defines the initial power-spectrum of
fluctuations stemming from the inflation period, and the power is
normalised by the observed abundance of galaxy clusters yielding the
the present root-mean-square matter fluctuation averaged over a sphere
of radius $8\,h^{-1}\,$Mpc, $\sigma_8 \approx 0.81$, and the model has
no curvature.

\subsection{Tensions between the SMoC and observations}
\label{sec:SMoC_tensions}

The literature on the problems facing the SMoC comprises a large body
and the reader is encouraged to explore it and the suggested solutions
(e.g. \cite{Buchert+2018, DiValentino2022, DelPopolo+2022,
  Mavromatos2022, Melia2022b}). Here only a concise discussion is summarised of
some of the tensions between the SMoC and data with the aim of
spelling out the minimal challenges that any new cosmological model
(e.g. the AMoC, Sec.~\ref{sec:AMoC}) must address in order to qualify
as a tangible alternative model.

The need to introduce AH1 (inflation) and AH3 (dark energy) is met
with four fundamental tensions: {\bf (FT1)} Observational and
experimental data disfavour the inflationary scenarios at the core of
the SMoC. Ljjas, Steinhardt \& Loeb (2017, \cite{Ljjas+2013}) argue
that the theory of inflation needs to be questioned because the
simplest models of inflation have been ruled out by Planck data,
inflation is unstable and postmodern inflation lies outside of normal
science as it cannot be tested (see also \cite{Ljjias+2017}). {\bf
  (FT2)} The ``cosmological energy catastrophe'' \citep{Kroupa+2010}
arises from the fact that, in the SMoC, there is currently no accepted
standard as to why the existence of dark energy would not violate the
principle of energy-momentum conservation. Dark energy is taken to be
a constant vacuum energy density. So, as the universe expands, the
larger volumes imply that in the SMoC, the Universe tends towards an
infinite energy content \cite{Harrison1995, Peacock1999,Baryshev2008,
  Baryshev2015} -- i.e., cosmological models with dark energy do not
conserve energy.  In particle physics, dark energy can be interpreted
as the energy density of the vacuum. It has been argued (e.g., Zumino
1975 \citep{Zumino1975}) that if there exist equal numbers of degrees
of freedom for bosons and fermions, then the total vacuum energy
equals zero. If this symmetry were broken, then a field would have a
net sum for the zero-point energy of normal modes. The cut-off scale
of its momentum must be set at the Planck scale where the effects of
gravity and quantum mechanics become comparable. The resulting
prediction from the SMoPP exceeds the observed energy density of dark
energy by approximately 120 orders of magnitude. This discrepancy is
widely regarded as the worst prediction in the history of
physics. Hence, the dark energy term defies current understanding of
vacuum energy in the context of space expansion and the SMoPP.
(e.g. \cite{Weinberg1989, Baryshev+1994, Afshordi2012, Rafelski+2009,
  Burgess2013}).  {\bf (FT3)} The large-scale properties implied by
dark matter would exclude any candidate particle (or
system)\footnote{A now disfavored dark matter candidate are massive
  compact halo objects (MACHOs), which have been ruled out by
  microlensing surveys in the mass range $0.6 \times
  10^{-7}$~--~$15 M_{\odot}$ \citep{Tisserand+2007} and
  $0.3$~--~$30 M_{\odot}$ \citep{Alcock+2001}.} that originates from
the SMoPP.  There is no evidence for the existence of dark matter
particles that is independent of the assumed gravitational theory (A3
above), since all laboratory-based searches have lead to null results
\citep{Snowmass22}.  {\bf (FT4)} Another major unsolved problem of the
SMoC that is relevant for all alternative models is the baryon
asymmetry, because at the moment the physical process of creating more
matter than antimatter during inflationary baryo-leptogenesis remains
unknown (e.g. \cite{Sarkar+1999, Nikulin+2021, Mukaida+2022}).

The above FTs are not a reason to discard the model since they may
merely indicate our current lack of physical interpretation of
inflation and dark energy. Upholding the SMoC to be the correct model
necessitates a continued significant research effort to progress on
these tensions (e.g. \cite{Lahav2023}).  The lack of experimental
evidence for dark matter particles is also not detrimental, because
the dark matter particles may have the property that the non-gravitational
interaction cross section with SMoPP constituents may be negligible,
rendering the dark matter particles undetectable in any experiment. It is
therefore permissible to assume the SMoC to be valid.

As any mature theory, the SMoC needs to be tested and compared with
advancing observational data. In the past, the SMoC (in the form of
the $\Lambda$CDM model) has been heralded as being successful in
accounting for observational data (e.g. \cite{Primack2012,
  Blanchard+2022}) while serious issues had also been noticed
\citep{Kroupa+2010, Kroupa2012, Kroupa2015}.  At the present time, the
comparison of the SMoC with state-of-the art observations has lead to
the recognition of certain tensions:

{\bf (T1)} {\it Lack of observational evidence for dynamical
  dissipation by extended particle-based dark matter halos}: If dark
matter particles do exist, then they will populate massive and
extended dark matter halos around the baryonic galaxy components
(e.g. \cite{Stewart+2008,Sales+2017, Moster+2020}).  This is a strict
prediction of the SMoC.  Galaxy--galaxy encounters will consequently
suffer dynamical dissipation (i.e. Chandrasekhar dynamical friction
when one galaxy is significantly minor compared to the other) when the
extended dark matter halos begin to touch each other and the galaxies
will slow down their relative velocity and fall towards each other to
merge.  This dissipation can be calculated
(e.g. \cite{BinneyTremaine1987, BinneyTremaine2008}). It leads to
every galaxy growing, in the SMoC, through many mergers to its
present-day mass. Likewise, the bars of disk galaxies are rigid
rotators and as a body suffer dynamical dissipation on the dark matter
halos of their galaxies -- in each barred galaxy the dark matter halo
absorbs the kinetic energy of the rotating bar which consequently
slows down.  In addition to the lack of evidence in a variety of data
for mergers dominating galaxy growth noted previously
(\cite{Kroupa2015} and references therein), this dynamical dissipation
is excluded by the observed fast rotation speeds of galactic bars
which should typically be slow in the presence of the dark matter
halos (Roshan et al. 2021, \cite{Roshan+2021}). The calculations
falsify the existence of dark matter halos with a confidence of
$10\,\sigma$.  That is, even when using distinct cosmological
structure formation codes, SMoC hydrodynamical simulations are unable
to replicate the bar pattern speeds observed in real galaxies.

An interesting result concerning dynamical dissipation due to the
expansive and massive dark matter halos of the SMoC are available for
the nearby ($\approx 3.3\,$Mpc distant) M81~group of galaxies. The
distribution of HI gas throughout the system of more than three
interacting galaxies constrains the past encounters between the
galaxies, which, in the SMoC, are all together within the inner region
of their dark matter halos such that dynamical dissipation is
pronounced.  Yun (1999 \cite{Yun1999}) reports that the model system
merges too rapidly to account for the observed tidal structures. This
conclusion is shared by the independent numerical modelling by Thomson
et al. (1999 \cite{Thomson+1999}). Both of these groups thus come to
the same conclusion, namely, that numerical modelling with dark matter
halos does not lead to agreement with the observed system. Both of
these are conference proceedings and neither group returned to the
problem.  The more recent modelling of the M81~system by Oehm et
al. (2017 \cite{Oehm+2017}) verifies the absence of dark-matter-based
solutions. This work shows that the three major galaxies in the system
must have approached each other from large (Mpc-scale) distances, in
violation of the Hubble flow, to all meet at virtually the same time
(the present time when we observe the system) just before they
merge. This is arbitrarily unlikely in the SMoC, leading to the strong
exclusion of the existence of dark matter halos.

\begin{figure*}[ht!]
\centering
\includegraphics[width=0.7\linewidth]{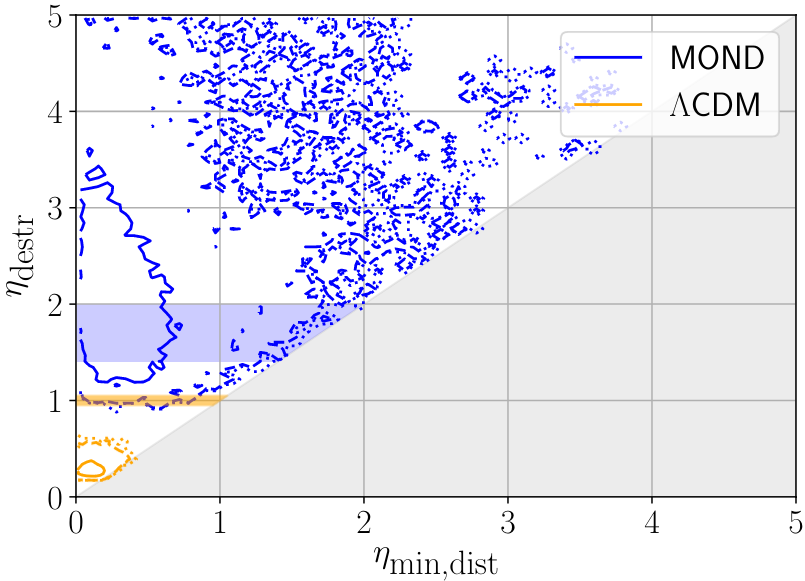}
\caption{The tidal susceptibility,
  $\eta_{\rm rtid} = r_{\rm h} / r_{\rm tid}$, of dwarf galaxies in
  the Fornax galaxy cluster. Dwarf galaxies in dark matter halos have
  a large tidal radius, $r_{\rm tid}$, compared to their observed
  half-mass radius, $r_{\rm h}$, and thus a small $\eta_{\rm rtid}$
  value. A dwarf will be visibly tidally perturbed (i.e. close to
  destruction) if $\eta_{\rm destr}\approx 1$ (the thick horizontal
  orange bar), while the observed galaxies have, when interpreted in
  terms of the SMoC being valid, $\eta_{\rm rtid}\approx 0.25$ (the
  orange contours).  The observed degree of tidal perturbation
  together with the expected dark matter halo masses places the
  $\Lambda$CDM models well below the nominal
  $\eta_{\rm destr}\approx1$ value, therewith invalidating the
  dark-matter halo hypothesis at more than $5\,\sigma$
  confidence. Without dark matter and in MOND (see
  Appendix~\ref{sec:MOND}), on the other hand, the expected
  $\eta_{\rm rtid}$ values are comfortably in the region where the
  dwarf galaxies are expected to be tidally perturbed (blue shaded
  region).  We show the $1\,\sigma$ (inner solid line), $3\,\sigma$
  (dashed line), and $5\,\sigma$ (outer dotted line) confidence region
  for MOND (blue) and $\Lambda$CDM (orange) models.  The grey-shaded
  region is unphysical because for a dwarf within this region the
  threshold at which it starts to appear perturbed would be higher
  than the threshold at which the dwarf is destroyed by the tides. For
  more details see fig.~14 in \citep{Asencio+2022}. With kind
  permission from Asencio et al. (2022 \cite{Asencio+2022}).  }
\label{fig:Fornaxdwarfs}
\end{figure*}

{\it Independently of dynamical friction}, a dark matter halo protects
a dwarf galaxy from being tidally perturbed for most of its orbital
history before it merges or disintegrates
\cite{Kazantzidis+2004}. There is thus only a brief window of
opportunity to observe significantly perturbed baryonic galaxies.  The
large number of tidally deformed dwarf galaxies in the Fornax galaxy
cluster rules this out, as shown in the detailed study by Asencio et
al. (2022 \cite{Asencio+2022}, Fig.~\ref{fig:Fornaxdwarfs} here).
That the ultra-diffuse dwarf galaxies (UDGs), some of which are
reported to be lacking dark matter, are in strong disagreement with
galaxies formed in the SMoC has been reported by Haslbauer et
al. (2019a \cite{Haslbauer+2019a}). And, additional significant
evidence against the existence of CDM and WDM particles comes from the
very recently reported smooth lensing signal of multiply lensed images
of background galaxies by Amruth et al. (2023
\cite{Amruth+2023}). With the authors knowing dark matter exists, the
lack of lensing sub-structure is explained with axions with a de
Broglie wavelength of $\lambda_{\rm dB}\approx 180\,$pc. This however
poses problems for the presumed dark matter content of ultra faint
dwarf galaxies (UFDs) \cite{Simon2019} with radii that are smaller
than $\lambda_{\rm dB}$.  Dalal \& Kravtsov (2022
\cite{DalalKravtsov2022}) conclude that the constraints provided by
the UFDs make fuzzy dark matter have indistinguishable properties to
cold dark matter on the scales probed.  Neither this nor the above
constraints on the existence of dark matter particles using dynamical
friction and Fornax dwarf galaxies are discussed by Amruth et
al. (2023 \cite{Amruth+2023}). Also, the reader is reminded of an
early observation by Dubinski et al. (1995 \cite{Dubinski+1996}): the
great lengths of the tidal tails produced during galaxy--galaxy
encounters are not possible with the massive and extended dark matter
halos that must surround major galaxies in the SMoC because the
tidally expelled material cannot propagate out of the deep potential
wells to the observed distances from the host galaxies.

{\it Summarising}, the sum of the above evidence using completely
different and independent methods compellingly disfavours the
existence of dark matter particles of any mass.

In addition to this failure of the dark-matter models to match the data, additional tensions have emerged:

{\bf (T2)} {\it Asymmetrical tidal tails}: Open star clusters that
orbit the Galaxy on near-circular trajectories lose stars through the
energy equipartition process. As a result, the evaporation of stars
into the leading or trailing tails in a cluster in the Solar
neighbourhood is a stochastic process, as shown by Pflamm-Altenburg et
al. (2023 \citep{Pflamm-Altenburg+2023}). That is, within Poissonian
differences, there should be an equal number of stars in both
tails. But the tidal tails of open star clusters in the Solar
neighbourhood are asymmetrical with the nearest cluster, the Hyades,
at a distance of about~45~pc showing more stars in its leading tail at
the $6.5\,\sigma$ confidence level (Kroupa et al. 2022
\citep{Kroupa+2022}). These data rule out the validity of Newtonian
gravitation on the~pc scale unless another explanation for this
asymmetry can be found. The only possibility would be that the Hyades
suffered an encounter with a massive ($\approx 10^7\,M_\odot$)
perturbing object, as demonstrated by Jerabkova et al. (2021
\cite{Jerabkova+2021}). But there is no independent evidence for such
a massive perturber in the Solar neighbourhood. Also, the detailed
structure of the observed tidal tails of the Hyades shows a radial
symmetry between the leading and trailing tail despite the different
number of stars in them, which is not plausible if the number
asymmetry had been produced by an encounter
\cite{Pflamm-Altenburg+2023}.

{\bf (T3)} {\it Exponential galactic disks and rotation curves}: The
merger-driven dark-matter model has not yielded an explanation as to
why observed disk galaxies have radial exponential surface mass
density profiles (Wittenburg et al. 2020 \citep{Wittenburg+2020}).
Given the observed distribution of baryonic matter in a disk galaxy,
it is not possible to calculate or predict from this the rotation
curve of the galaxy because the model dark matter halos show a large
dispersion of concentrations and shapes and spins
(e.g. \cite{Moster+2020}). Given the dispersion of concentrations and
masses, the model rotation curves do not fit the observed shapes
\citep{McGaugh+2007, McGaugh2015}. This mismatch suggests the observed
rotation curves to not be related to dark matter halos. Small-scale
($\simless 1\,$kpc) wiggles in the rotation curves are matched by
density variations in the baryonic matter distribution and cannot be
modelled with dark-matter halos (``Renzo's Rule'',
e.g. \cite{FamaeyMcGaugh2012}).

{\bf (T4a)} {\it The Impossible Local Group: Disks of
  Satellites~/~Planes of Satellites}: The Galaxy
\citep{LyndenBell1976, Kroupa+2005, Metz+2008, PawlowskiKroupa2020},
Andromeda \citep{Koch+2006, Metz+2007, Ibata+2013}, Centaurus~A
\citep{Tully+2015, Mueller+2021}, M81 \citep{Chiboukas+2013}, NGC~253
\citep{Martinez+2021} and the galaxy pair NGC~4490/85 (Karachentsev \&
Kroupa, submitted) and most other galaxies (Ibata et al. 2014
\cite{Ibata+2014}) have a large fraction of their satellite galaxies
arranged in vast rotating planar structures.  Since the orbits of the
satellites are confined to disks, these systems are reminiscent of
planetary systems requiring a dissipational process for their
formation rather than the spheroidal distribution of dark-matter
bearing satellite galaxies expected from the merger tree in the SMoC
(Pawlowski 2021a,b \cite{Pawlowski2021a, Pawlowski2021b}).  Even on
its own, the Milky Way disk of satellites presents a significant
challenge for the SMoC as pointed out for the first time by Kroupa et
al. (2005 \cite{Kroupa+2005}, update \cite{PawlowskiKroupa2020}).  The
combined tension of only three (the Galaxy, Andromeda and Centaurus~A)
existing simultaneously is above the 5~sigma confidence (Asencio et
al. 2022, \cite{Asencio+2022}). But in total there are now six such
systems, and in fact, as pointed out by Chiboukas et al.  ("in the few
instances around nearby major galaxies where we have information, in
every case there is evidence that gas-poor companions lie in flattened
distributions"), the existence of flattened, rotating systems of
satellite galaxies appears to be the norm rather than the exception.
That is, the dark-matter-based structure formation models are
inconsistent with the observations with more than 5~sigma
confidence.\footnote{ Recently, some authors have revisited the Disk
  of Satellite (DoS) problem arguing that the Milky Way satellite
  system is short-lived but normal in $\Lambda$CDM: Xu et al. (2023
  \cite{Xu+2023}) find the probability of the observed configuration
  of 11~classical satellites to be about $1/231$.  Ignoring that the
  Milky Way, Andromeda and Cen~A actually have more satellite galaxies
  in their disk configurations than only~11 in each, observing three
  such independent plane-of-satellite systems simultaneously in our
  immediate cosmological vicinity would have a probability of
  $8\times 10^{-8}$ (consistent with the more accurate calculation
  available in Asencio et al. 2022 \cite{Asencio+2022}). Despite this,
  the $\Lambda$CDM authors argue that the DoS is consistent with the
  SMoC, therewith apparently introducing a renormalisation of the
  concept of probability. Sawala et al. (2022 \cite{Sawala+2022})
  likewise renormalise the likelihood of the Milky Way DoS elevating
  it to the status of being a normal occurrence, finding the DoS of
  11~classical satellite galaxies to be a short-lived configuration
  but count this to be consistent with the SMoC. Since the
  configuration lasts a few hundred~Myr only, it would constitute a
  remarkable coincidence of observing the DoS just now when
  higher-life appeared on Earth: it's emergence being heralded by the
  Great Constellation of Satellite Galaxies. However, they
  \cite{Sawala+2022} base their analysis on erroneous calculations
  (e.g. the apogalactic distances of the Large and Small Magellanic
  Clouds are far too small and the errors on proper motion
  measurements of the satellite galaxies are taken inappropriately
  into account in the orbit integration causing the unnatural
  short-lived phase of the DoS) and misrepresent the previous work by
  \cite{PawlowskiKroupa2020} in terms of the combination of Gaia and
  HST proper motions. \cite{Kanehisa+2023} confirm that a large
  fraction of the Cen~A satellites are in a coherently rotating disk.}
The only known physically plausible process to create the observed
phase-space correlated satellite galaxies is through them having
formed as tidal dwarf galaxies in tidal arms pulled out in
galaxy--galaxy encounters. Such tidal dwarf galaxies lack dark matter
though and thus the SMoC predicts there to be primordial dwarf
galaxies (being dominated by dark matter) and tidal dwarf galaxies
(lacking dark matter) which therefore have different radii
\cite{Kroupa2012}. This {\it dual dwarf galaxy theorem} is evident in
high-resolution SMoC simulations (Haslbauer et al. 2012b
\cite{Haslbauer+2019b}), who show that the observed population of both
types of dwarf galaxies (being indistinguishable
\cite{DabringhasuenKroupa2013}) poses a significant tension for the
SMoC.

\begin{figure*}[ht!]
\centering
\includegraphics[width=\columnwidth]{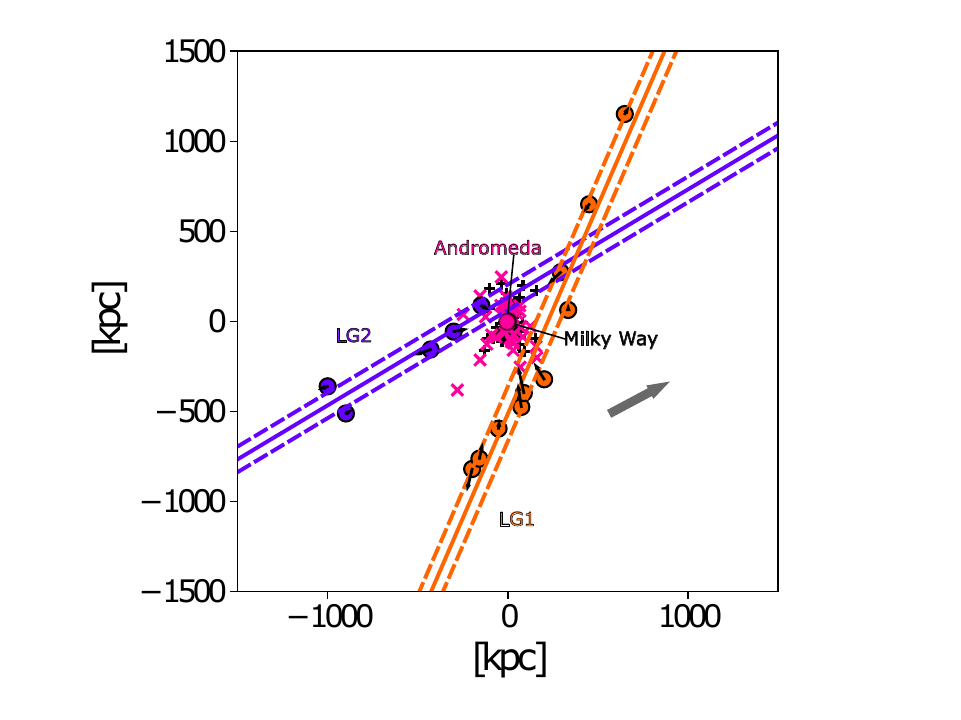}
\caption{The 
highly symmetrical multiple-planar
 arrangement of matter in the Local Group (LG) of galaxies. The observer is at infinity and looking at the LG along the direction joining the Milky Way (MW, black) and Andromeda (M31, magenta). The MW satellite galaxies (with Galactocentric distance $\,\le300\,$kpc, most being of dwarf-spheroidal, dSph, type) are the black $+$ signs, while the M31 satellites (most being of dSph type, with M31-centric distance $\,\le300\,$kpc, except for the Pegasus dwarf-irregular (dIrr) and Andromeda~XVI, both of which are further but are considered to belong to the Great Plane of Andromeda, i.e., to M31's DoS, because of their positions and velocity vectors) are magenta crosses. All other galaxies (most being of dIrr type) in the LG are shown as orange and purple circles, being $d_{\rm LG}\, \le 1500\,$kpc from the point midway between the MW and M31. These galaxies are in two major planes indicated by the purple (the "LG2" plane) and orange (the "LG1" plane) solid lines. These planes are about 100~kpc thick (root-mean-square heights $\,\approx50\,$kpc, the dashed lines), and are approximately equidistant from the line joining the MW and M31 (the viewing direction here). No galaxy or system has been detected outside of these planes.
 The black solid arrows show the Galactocentric line-of-sight velocities of the galaxies (as seen from the Sun), and the large grey arrow shows the direction of the motion of the Local Group w.r.t. the CMB.   Updated version of fig.~9 in \cite{Pawlowski+2013} (using data from \cite{McConnachie2012}, updated until~2021). 
}
\label{fig:LoGr1}
\end{figure*}

{\bf (T4b)} {\it The Impossible Local Group: it's highly symmetrical multiple-planar
  structure}: Pawllowski, Kroupa \& Jerjen (2013, \cite{Pawlowski+2013}) discovered that all non-satellite
galaxies in the Local Group are situated in two $\approx 100\,$kpc
thick Mpc-scale planes (the {\it dominant plane}, LG1, containing more
galaxies), as shown in Fig.~\ref{fig:LoGr1}.
In Fig.~\ref{fig:LoGr1}, the line joining the Milky Way and Andromeda
is perpendicular to the plane of the figure, and the Local Group moves
w.r.t. the CMB along the upper right diagonal.  The two planes, LG1
and LG2, are also perpendicular to the plane of the figure (both
planes being seen edge-on) and are symmetrically and approximately
equidistantly placed about the line joining Andromeda and the Galaxy.
The origin of this Mpc-scale highly symmetrical multiple-planar
structure is entirely unknown. Why do the galaxies not fall towards
the barycentre of the Local Group? -- the Local Group should be
virialising, and in this process becoming pressure supported and
spheroidal.

Just outside of the Local Group is another similar (i.e. thickness of
$\approx 100\,$kpc) Mpc-scale plane of 5~galaxies, the NGC$\,$3109
association, which is parallel to and offset by $0.3-0.5\,$Mpc from
the dominant plane LG1 (Pawlowski \& McGaugh 2014,
\cite{PawlowskiMcGaugh2014}).  A few Gyr ago the NGC~3109 association,
itself a plane approximately parallel and similar to LG1 in extension
and thickness, was in the Local Group as it coherently moves away from
it. The physical origin of this plane and the reason for its
orientation and apparent association and movement relative to the
dominant LG1 plane defy current understanding.

These data that uncover a rather extreme symmetrical organisation of
matter over Mpc scales demonstrate that an infall-history in the sense
of the SMoC in which groups form from mergers is clearly ruled
out. The placement and motion away from the Local Group of the
NGC~3109 association also rules out the action of Chandrasekhar
dynamical friction on dark matter halos, because the coherent receding
motion of the entire NGC~3109 association constituting galaxies with
significantly different masses would not be possible under the action
of dynamical friction, this conclusion being consistent with~T1
above. Are other nearby galaxy groups also as organised? This is
unclear because the 3D relative distances between galaxies in a group
are too large for the detection of a Local-Group-like symmetric 3D
structure.

Thus, both the highly symmetrical multiple-planar structure and the
motion of the NGC~3109 association falsify the SMoC with essentially
infinite confidence.  At the moment there exists no theoretical
framework able to explain these data.

\begin{figure*}[ht!]
\centering
\includegraphics[width=0.7\linewidth]{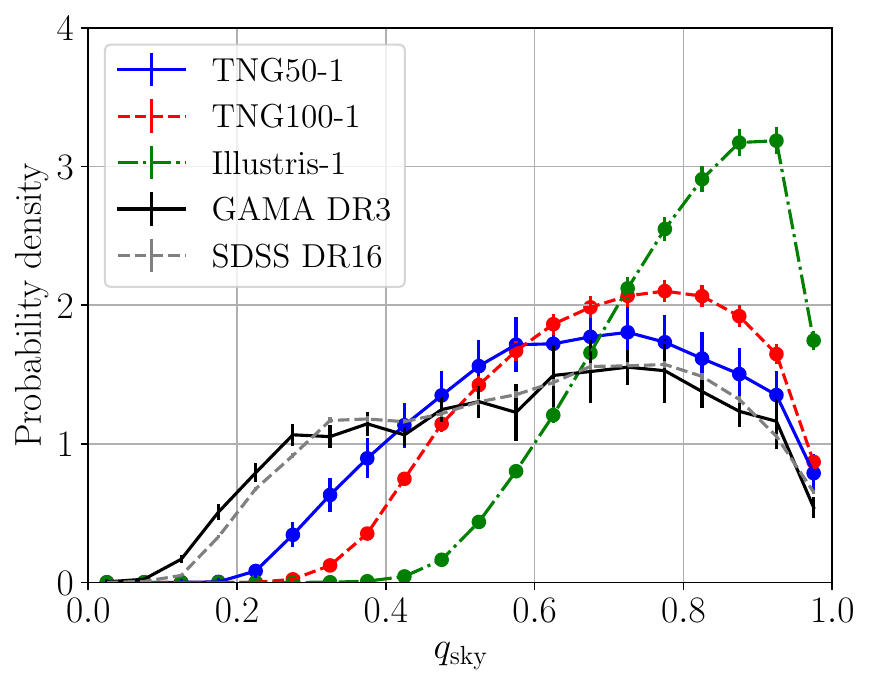}
\caption{The probability density distribution of the on-sky
  ellipticity, $q_{\rm sky}$, of galaxy images in the real Universe
  (solid black and dashed grey lines) in comparison to SMoC
  simulations (green, red and blue data, EAGLE simulations yielding
  comparable distributions, see fig.~4 in
  \cite{Haslbauer+2022}). Perfectly thin disk galaxies without
  observational uncertainties give a distribution between
  $0.1<q_{\rm sky}<0.5$ only (fig.~14 in \cite{Haslbauer+2022}), while
  spherical galaxies have $q_{\rm sky}=1$. The SMoC models are in more
  than $13\,\sigma$ disagreement with the data (table~3 in
  \cite{Haslbauer+2022}). Reproduced with permission from Haslbauer et al., being 
  fig.~4 in \cite{Haslbauer+2022}.  }
\label{fig:morph}
\end{figure*}

{\bf (T5)} {\it Only a few per cent of all galaxies are
  ellipticals. These formed on the short downsizing time-scale, and
  the formation of super-massive black holes (SMBHs) likewise occurred
  extremely early and rapidly}: It is known by observation since more
than three decades that elliptical galaxies make up only a few~per
cent of all galaxies with the vast rest being rotationally supported
disk galaxies \cite{Binggeli+1988, Delgado+2010}.  In 2014
Vogelsberger et al.  \cite{Vogelsberger+2014b} wrote ``{\it Simulating
  the formation of realistic disk galaxies, like our own Milky Way,
  has remained an unsolved problem for more than two decades. The
  culprit was an angular momentum deficit leading to too high central
  concentrations, overly massive bulges and unrealistic rotation
  curves. The fact that our calculation naturally produces a
  morphological mix of realistic disk galaxies coexisting with a
  population of ellipticals resolves this long-standing issue. It also
  shows that previous futile attempts to achieve this were not due to
  an inherent flaw of the $\Lambda$CDM paradigm, but rather due to
  limitations of numerical algorithms and physical modelling.}''  This
is interesting in view of Stewart et al. having written in~2008
\cite{Stewart+2008},``{\it Our results raise serious concerns about
  the survival of thin-disk-dominated galaxies within the current
  paradigm for galaxy formation in a $\Lambda$CDM universe. In order
  to achieve a ~70~\% disk-dominated fraction in Milky Way-sized
  $\Lambda$CDM halos, mergers involving
  $\approx 2 \times 10^{11}h^{-1}\,M_\odot$ objects must not destroy
  disks. Considering that most thick disks and bulges contain old
  stellar populations, the situation is even more restrictive: these
  mergers must not heat disks or drive gas into their centers to
  create young bulges.}''  In spite of the positive depiction of the
SMoC results by Vogelsberger, Haslbauer et al. confirm, in 2022
\cite{Haslbauer+2022}, the conclusion by Stewart et al. by
demonstrating that the most modern and highest resolution
hydrodynamical SMoC simulations are in more than $13\,\sigma$ conflict
with the above observation that disk galaxies dominate by far the
galaxy population (Fig.~\ref{fig:morph}).

This severe conflict between the real galaxy population and the SMoC
calculations is robust because it persists in the most recent highest
resolution simulations that involve very different types of computer
codes and sub-grid baryonic physics descriptions. Haslbauer et
al. (2022 \cite{Haslbauer+2022}) demonstrate that even a massive
improvement in resolution will not be able to remedy this disaster.
The reason for this model failure is that galaxies grow mostly through
mergers and these destroy the existing thin disks such that the
majority of galaxies formed in SMoC simulations are puffed-up,
bulge-dominated or spheroidal. In contrast to the model, there is a
pre-prominence of very thin disk galaxies in the real Universe, a
significant fraction (roughly 30~per cent) lacking bulges
\citep{Kormendy+2010}.

While elliptical galaxies form readily in the SMoC, Eappen et
al. (2022 \cite{Eappen+2022}) demonstrate that the age distribution of
the stellar particles in these is much broader than in real observed
ellipticals which are known to follow downsizing such that the more
massive galaxies formed faster and earlier (\cite{Yan+2021,
  Salvador+2022} and references therein). The work by Eappen et
al. shows that the distribution of ages of stellar particles in the
model early-type galaxies is inconsistent at more than $5\,\sigma$
confidence with the short formation time-scales of the real early-type
galaxies.  In the SMoC elliptical galaxies need time to build-up from
many mergers.

A directly related problem for the SMoC is that there is no natural
physical process that would form SMBHs because the conditions in the
first dwarf dark matter halos cannot form these and mergers typically
eject any forming massive black holes if present \citep{Lena+2014,
  Kroupa+2020}. While simulations of galaxy formation in the SMoC need
to place pre-existing seed-SMBHs into the dark matter halos
(e.g. \cite{DeGrafSijacki2020}) without understanding where they come
from, a physically more accurate calculation of the rapid formation of
early-type galaxy components (ellipticals and classical bulges) shows
SMBHs to arise very naturally and without exotic physics at their
centres (Kroupa et al. 2020 \cite{Kroupa+2020}). However, in the SMoC
early-type galaxy components cannot form rapidly enough, as discussed
above.

{\bf (T6)}~{\it The Hubble Tension and the KBC void}: The measurement
of the Hubble-Lemaitre constant using supernovae~1a (SN1a) has become
highly significantly discrepant with the same constant deduced from
the observed CMB (if it is interpreted as being the standard CMB,
i.e. the surface of last scattering, SoLS, or the photosphere of the
Hot Big Bang, see Sec.~\ref{sec:conclusions}).  The SN1a measurement
of the current expansion rate of the Universe out to a few hundred~Mpc
yields the local value of the Hubble-Lemaitre constant,
$H_0^{\rm local} = 73.04\pm 1.04\,{\rm km}\,{\rm s}^{-1}\,{\rm
  Mpc}^{-1}$ \citep{Riess+2022}. The global expansion rate is measured
from the CMB to be
$H_0^{\rm global}= 67.4\pm0.5\,{\rm km}\,{\rm s}^{-1}\,{\rm Mpc}^{-1}$
\citep{Planck2018}. Both values confirm that the present-day Universe
is expanding more rapidly than one without dark energy if these
numbers are interpreted with the assumption that the SMoC be
valid. However, the difference between $H_0^{\rm local}$ and
$H_0^{\rm global}$ (the ``Hubble Tension'') is extremely significant.
The usual interpretation of the Hubble Tension is that it points
towards a more complicated physical theory for dark energy, such as
early dark energy models, that temporarily change the expansion rate
(\cite{Poulin+2023} for an overview) and thus also affecting the age
of the universe. Goldstein et al. (2023 \cite{Goldstein+2023}) rule
out canonical dark energy models formulated to solve the Hubble
Tension using the properties of absorption systems along the line of
sight.

\begin{figure*}[ht!]
\centering
\includegraphics[width=0.7\linewidth]{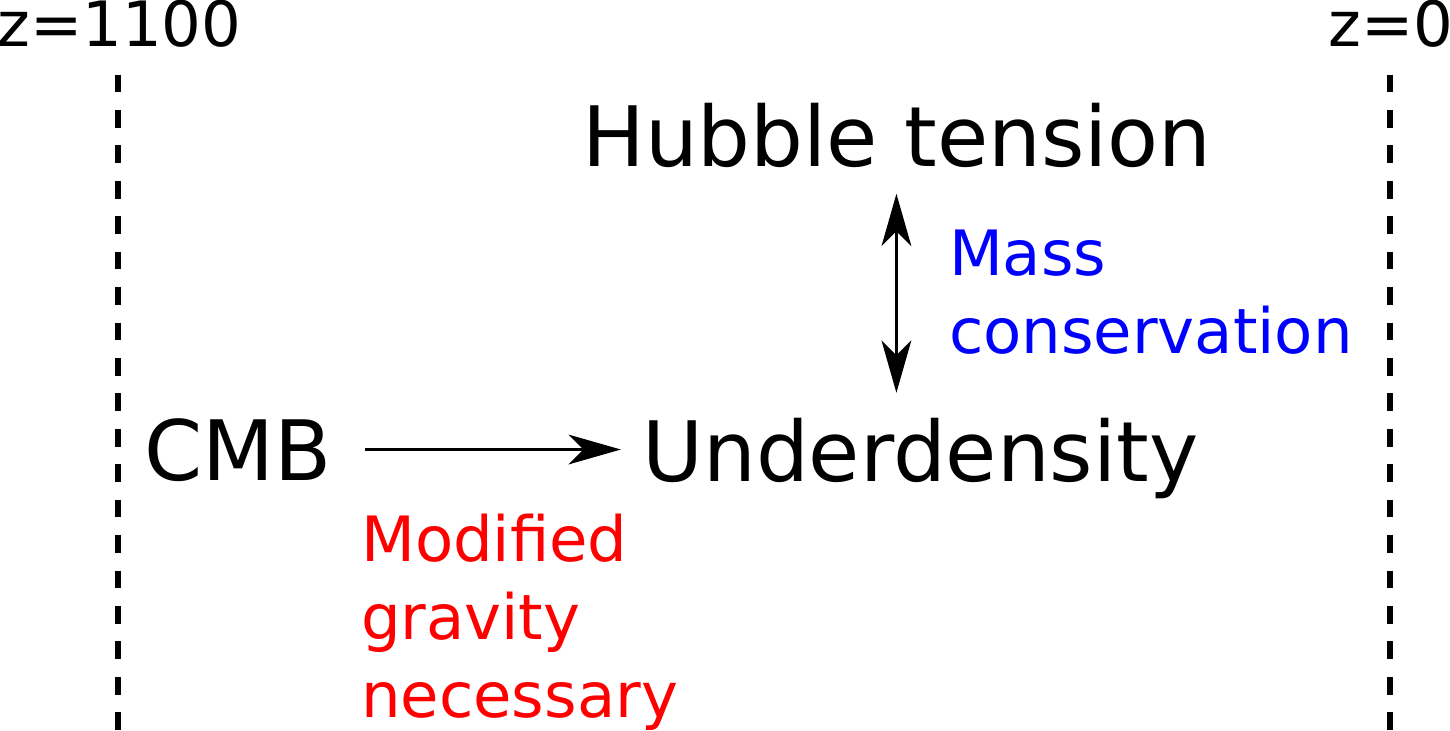}
\caption{Taking the CMB as the initial boundary condition at redshift
  $z=z_{\rm e}=1100$ (left vertical dashed line) which defines the
  allowed initial density variations in the cosmological model,
  structure growth in the model is calculated (black arrow). Assuming
  all matter to be created during the inflationary era (for
  consistency with the assumed CMB boundary condition) implies mass
  conservation and the calculation yields the possible density
  contrasts and void sizes at $z=z_{\rm e}=0$. Given that the SMoC
  models cannot account for the magnitude and extension of the
  observed matter density contrasts, new models are required that need
  effectively stronger gravitational forces to grow structures from
  the assumed CMB boundary condition and assuming matter conservation
  to remain valid.  An example of such a model is the AMoC, i.e. the $\nu$HDM model \cite{Haslbauer+2020}. Reproduced with permission from Haslbauer et al. (2020 \cite{Haslbauer+2020}, their fig.~11).  }
\label{fig:HT1}
\end{figure*}

\begin{figure*}[ht!]
\centering
\includegraphics[width=0.7\linewidth]{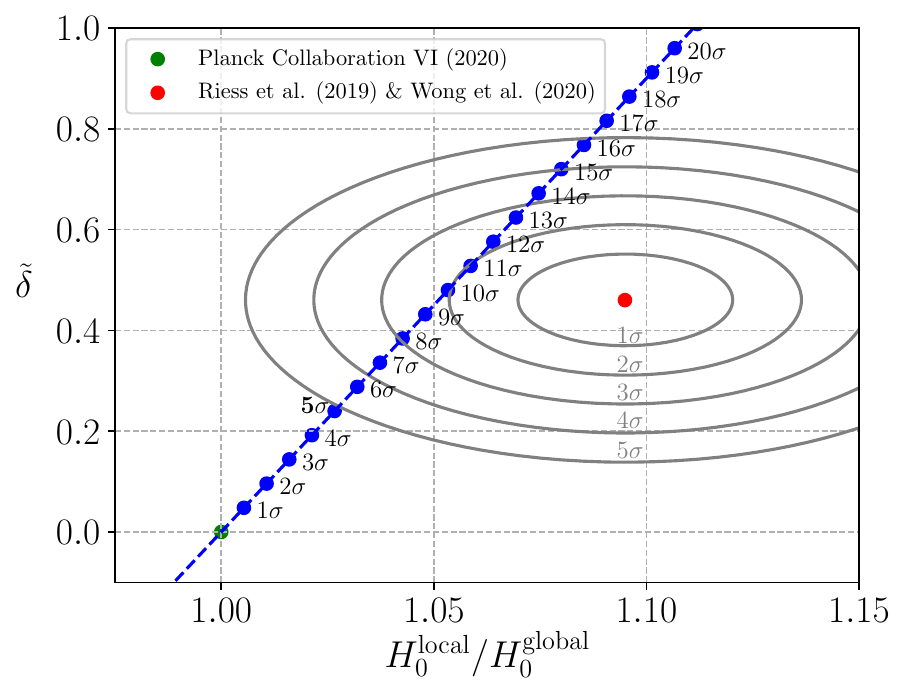}
\caption{The observed local under-density not corrected for the redshift space distortion (RSD), $\tilde{\delta} = 1-\rho/\rho_0$, versus the ratio
  $H_0^{\rm local}/H_0^{\rm global}$.  $\rho$ is the mass-density of
  matter between a radius of $\approx 40\,$Mpc and $\approx 300\,$Mpc
  from the Local Group and $\rho_0$ is the corresponding cosmic mean
  mass density.  The green point is for no Hubble Tension and no
  under-density, i.e., it corresponds to the $\Lambda$CDM expectation
  on the probed spatial scale.  The red point is the observed datum
  with error ellipses plotted in terms of the measurement confidence
  levels.  The blue points show the apparent cosmic variance in
  $\Lambda$CDM at the indicated confidence
  level. Notice that a $5\,\sigma$ fluctuation is not enough to get
  within $5\,\sigma$ of the local observations. Thus, the observed KBC
  void (the red dot) cannot occur in a $\Lambda$CDM universe. This is
  fig.~2 in \cite{Haslbauer+2020}. }
\label{fig:HT2}
\end{figure*}

On the other hand, the difference between $H_0^{\rm local}$ and
$H_0^{\rm global}$ is a natural consequence of the about
$0.6\,$Gpc-scale under-density of matter by a factor of about~two to
20~per cent (the Keenan-Barger-Cowie, KBC, void) within which the
Local Group is situated because galaxies fall towards the sides of the
void as a result of the gravitational field (Hoscheit \& Barger 2018
\cite{HoscheitBarger2018}, Haslbauer et al. 2020
\cite{Haslbauer+2020}). The existence of such a Local Hole has been
known since at least~2003 through the work of Tom Shanks and
collaborators \cite{Frith+2003, WhitburnShanks2016, Shanks+2019a,
  Shanks+2019b, Wong+2022} and was also noticed by Igor Karachentsev
(2012 \cite{Karachentsev2012}).  While the Hubble Tension can
naturally be explained by this under-density (i.e., there is no Hubble
Tension since there is a void or hole), the scale and depth of this
KBC void or Local Hole is in tension with the SMoC (see
Fig.~\ref{fig:HT1}) at more than 5~sigma confidence
(Fig.~\ref{fig:HT2}) since the SMoC universe becomes homogeneous on
scales larger than about 100~Mpc \cite{Haslbauer+2020}.

Haslbauer et al. (2020 \cite{Haslbauer+2020}) thus argue that the
observed KBC~void needs effectively stronger gravitation for this
underdensity to be able to develop over a Hubble time. Note the
consistency of the conclusions reached here with those above: that
stronger effective gravitation is required for growth of the observed
density difference (Haslbauer et al. 2020, \cite{Haslbauer+2020}) is
consistent with the exclusion of the existence of dark matter (T1)
which also implies the necessity for stronger effective gravitation
given the mass deficit observed in galaxies when assuming
Einsteinian/Newtonian gravitation.

The existence of a 200~Mpc-scale under-density in the supernova data
was initially suggested by {\'O} Colg{\'a}in (2019 \cite{Colgain2019})
and has been confirmed independently over~90~per cent of the sky
\citep{Wong+2022}. Such a KBC void or Local Hole was declared to be
incompatible with the supernova distance-redshift relation (Kenworthy
et al. 2019 \cite{Kenworthy+2019}, but see Haslbauer et al. 2020
\cite{Haslbauer+2020} who discuss their error), and Brout et al. (2022
\cite{Brout+2022}) argue that their upgradet Pantheon+ sample shows no
evidence of any low-redshift variation in the matter density. But
their fig.~16 is in actuality consistent with a significant increase
of the matter density with increasing redshift up to a value of~0.3.
Subsequent independent analyses have found that supernova data alone
actually reveal a preference for a local void \citep{Lukovic+2020,
  KP2020}. Also, according to the results of \cite{Castello+2022}, the
preferred void parameters are remarkably close to those of the KBC
void. This thus indicates a consistency between results obtained from
galaxy number counts and from supernovae.  Watkins et al. (2023
\cite{Watkins+2023}) quantify the bulk flow of matter of about
450~km/s within a distance of about 270~Mpc finding it to be
consistent with the SMoC with a probability of $2.1\times
10^{-6}$. Noteworthy is that a similar bulk flow would be evident in
the data by the Local Group being present in the KBC void (see fig.~8
in \cite{Haslbauer+2020}; Mazurenko et al., submitted).  There are
indications that the void extends to $0.5<z_{\rm e} \simless 2$
through the Hubble-Lemaitre constant decreasing systematically with
increasing redshift to the global Planck value (fig.~A1 in
\cite{Wong+2020}, fig.~5 in \cite{Millon+2020}, fig.~2 in
\cite{Krishnan+2020}, fig.~5 in \cite{Dainotti+2021, Dainotti+2022};
fig.~3 in \cite{Jia+2022}; fig.~6 in \cite{HuWang2023}).

Cosmology-model-independent measurements of the Hubble-Lemaitre
constant are needed to shed light on the apparently somewhat
nontransparent issue of cosmological expansion. Applying
reverberation-mapped quasars for cosmological constraints cannot
determine the Hubble constant independently from the normalization of
the observed broad-line region radius-luminosity relation. First, one
can assume the value from other measurements (see e.g. Zajacek et
al. 2021 \cite{Zajacek+2021}). Second, if the cosmological constraints
of the quasar sample are consistent with the better established
cosmological probe, one can merge the two. Cao et al. (2022,
\cite{Cao+2022}) analyzed reverberation-mapped CIV+MgII quasars
jointly with BAO+H($z_{\rm e}$) data and obtained the Hubble-Lemaitre
constant value of
$H_0^{\rm rev}= 69.15\pm1.77\,{\rm km}\,{\rm s}^{-1}\,{\rm Mpc}^{-1}$
for the flat LCDM model.  The tight UV- and X-ray luminosity scaling
relation of quasars (Lusso et al. 2019 \cite{Lusso+2019}, 2020
\cite{Lusso+2020}) allow an independent method for measuring the
expansion rate of the Universe, but the results suggest a
significantly larger value of $H_0^{\rm UV-X}$ than $H_0^{\rm local}$
for $z_{\rm e}>3$ than is compatible with any of the other
measurements.  Apart from the "Hubble Tension" above, these disparate
measurements are for now not taken as being problematic for any
cosmological model, since understanding the baryonic processes that
lead to the reverberation process and the UV- and X-ray fluxes need to
be improved significantly.

A model-independent constraint on the expansion rate and thus age of
the Universe can however be obtained using cosmic chronometers and
from the oldest stars (see Sec.~\ref{sec:conclusions}).

{\bf (T7)}~{\it Large-scale matter inhomogeneities}: More than 5~sigma
evidence for large-scale ($>0.5\,$Gpc) inhomogeneities in the observed
matter distribution comes from line-of-sight motions of galaxy
clusters (Migkas et al. 2021 \citep{Migkas+2021}), the distribution
and line-of-sight velocities of quasars and active galactic nuclei
\citep{Secrest+2022}, as well as in the distribution of gamma-ray
bursts \citep{Horvath+2020}. Dupuy \& Courtois (2023,
\cite{DupuyCourtois2023}) use the CosmicFlows-4 catalogue to infer,
from the observed galaxy kinematical field, the variations of the
gravitational potential on~Mpc scales within a distance of about
$0.4\,$Gpc. They find that the inferred super-clusters are
significantly larger in volume than allowed by the SMoC, indicating
large-scale variations of the matter density that is not consistent
with the SMoC. The about~500 galaxies in the Local Cosmological Volume
(within radius around us of $\approx 11\,$Mpc) have good estimates of
their average and current star-formation rates, both being
approximately equal \cite{Kroupa+2020b}. Using this constraint,
Haslbauer et al. (2023, \cite{Haslbauer+2023} calculate the
star-formation rate density backwards in time finding that the
observed Lilly-Madau maximum of this quantity near a redshift of
about~1.9 (at "cosmic noon") is not evident in the Local Cosmological
Volume. Since the cosmological principle dictates the galaxies in the
Local Cosmological Volume to be representative of galaxies elsewhere,
Haslbauer et al. \cite{Haslbauer+2023} deduce that the maximum of the
star-formation rate density at "cosmic-noon" is due to a
$\approx 5.1\,$Gpc-distant matter over-density. This inhomogeneity,
observed by us at the present time, amounts to the Local Cosmological
Volume being a under-dense region by a factor of a few relative to the
regions observed about~5.1~Gpc in the past.

This evidence for large-scale inhomogeneity stands in contrast to the indirect evidence for a
smoother Universe gleaned from weak lensing analysis
(e.g. \cite{DiValentino2021c, Abbott+2023}).  This is related to the
``sigma~8 tension'' that comes from the analysis of CMB data yielding
a more clumpy universe than galaxy-cluster observations and may be
related to us being in the KBC void (see also ``pT4'' below). The
recent review by Aluri et al. (2022 \cite{Aluri+2022}) covers the above and further
observational indications for a violation of the cosmological
principle, these authors stating that ``{\it it is equally plausible that
precision cosmology may have outgrown the FLRW paradigm, an extremely
pragmatic but non-fundamental symmetry assumption.}''  Related to the
large-scale structure, (Mohayaee et al. 2021 \cite{Mohayaee+2021}) document that the CMB
reference frame is inconsistent with the AGN and quasar frame such
that corrections for peculiar velocities of the SN1a data would
invalidate, with high (but not quite $5\,\sigma$) confidence the
deduction that dark energy exists. Essentially, since the SN1a events
occur in galaxies that share bulk flows that systematically deviate
from the Hubble flow over large scales, inference on cosmological
expansion based on SN1a data does not yield the correct value of the
Hubble-Lemaitre constant.

Meanwhile many groups have found the Hubble-Lemaitre constant to
decrease with redshift (see T6 above).  This may be interpreted to be
due to a changing equation of state of the Universe (Krishnan et
al. 2021 \cite{Krishnan+2021}), but a plausible interpretation is also
that we are in a 5-Gpc (in radial distance) under-density within which
the KBC void or Local Hole is merely the small-scale (Gpc-scale
across) local observation causing the Hubble Tension. Either way, the
SMoC is invalidated by these observations and new physics is needed to
explain them.

\begin{figure*}[ht!]
\centering
\includegraphics[width=0.7\linewidth]{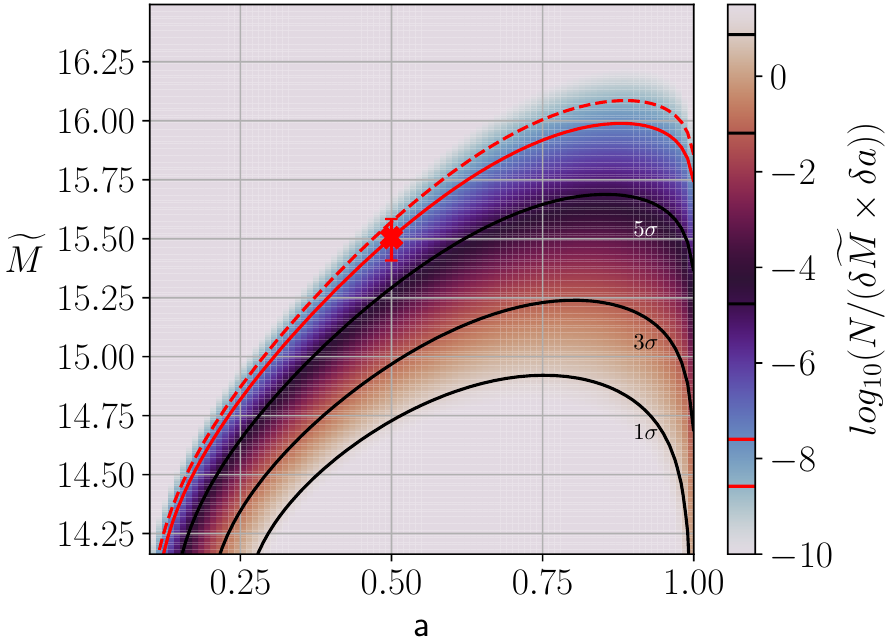}
\caption{The tension of observing an El Gordo-type cluster collision
  in the $\Lambda$CDM model with mass
  $\approx 10^{15.5}\,M_\odot = 10^{\widetilde{M}}\,M_\odot$, when the
  redshift $z_{\rm e} =1$ (expansion factor $a(t)=0.5$),
  is~$6.16\,\sigma$ (solid red contour line). The observed double
  cluster is at the red dot with 20~per cent uncertainty in mass.
  The levels of all the contours are shown on the colour bar. The
  $5\,\sigma$ contour is at $\widetilde{M}=15.293$ when
  $a(t)=0.5$. The El Gordo double galaxy cluster cannot exist in the
  SMoC.  For further details see fig.~7 in \cite{Asencio+2021}. }
\label{fig:ElGordo}
\end{figure*}

{\bf (T8)}~{\it The El Gordo and Bullet galaxy clusters}: The mass of
the binary galaxy cluster El Gordo
($\approx 2\times 10^{15}\,M_\odot$) and its existence so early in the
Universe (at redshift $z_{\rm e}=0.87$) is in tension with the SMoC at
$6.16\,\sigma$ confidence (Fig.~\ref{fig:ElGordo}), as shown by
Asencio et al. (2021 \cite{Asencio+2021}, update:
\cite{Asencio+23}). By $z_{\rm e}=0.87$ the SMoC model universe is not
old enough to have grown two neighbouring galaxy cluster that each
weigh near $10^{15}\,M_\odot$ and which have already fallen through
each other. The binary Bullet galaxy cluster adds to the tension as it
is also unlikely to occur in the SMoC \citep{KraljicSarkar2015}, both
binary clusters raising the
tension with the SMoC to $6.43\,\sigma$ \cite{Asencio+2021}.\\

\noindent In the following possible tensions are listed:\footnote{A ``possible
  tension'' is meant to be a tension that would falsify the SMoC with
  at least 5~sigma confidence provided further observations confirm
  the tension.}

Possible tension~1 {\bf (pT1)} arises in the {\it anomalies of the observed CMB} in view
of the physics of the inflationary epoch: \cite{Schwarz+2016} report
``{\it the lack of variance and correlation on the largest angular scales,
alignment of the lowest multipole moments with one another and with
the motion and geometry of the solar system, a hemispherical power
asymmetry or dipolar power modulation, a preference for odd parity
modes and an unexpectedly large cold spot in the Southern
hemisphere}''.

{\bf (pT2)} {\it Cosmological-scale hemispherical asymmetries}: The
southern celestial hemisphere has a more than $5\,\sigma$ larger number of
early-type galaxies with suppressed star formation rates than the
northern hemisphere (Javanmardi \& Kroupa 2017 \cite{JavanmardiKroupa2017}). While this was
interpreted to be due to unknown biases in the galaxy catalogues, the
southern celestial hemisphere also has more power and a higher
temperature in the observed CMB than the northern hemisphere, the
Planck data confirming the WMAP measurements \cite{Eriksen+2004,
  Schwarz+2016}. Furthermore, Shamir (2022 \cite{Shamir2022}) and Shamir \& McAdams (2022 \cite{ShamirMcAdam2022}) point out the existence of 
  correlations of disk galaxy spins over large regions of spatial volume with significant differences between the two hemispheres. 
  Are these asymmetries connected?

\begin{figure*}[ht!]
\centering
\includegraphics[width=0.7\linewidth]{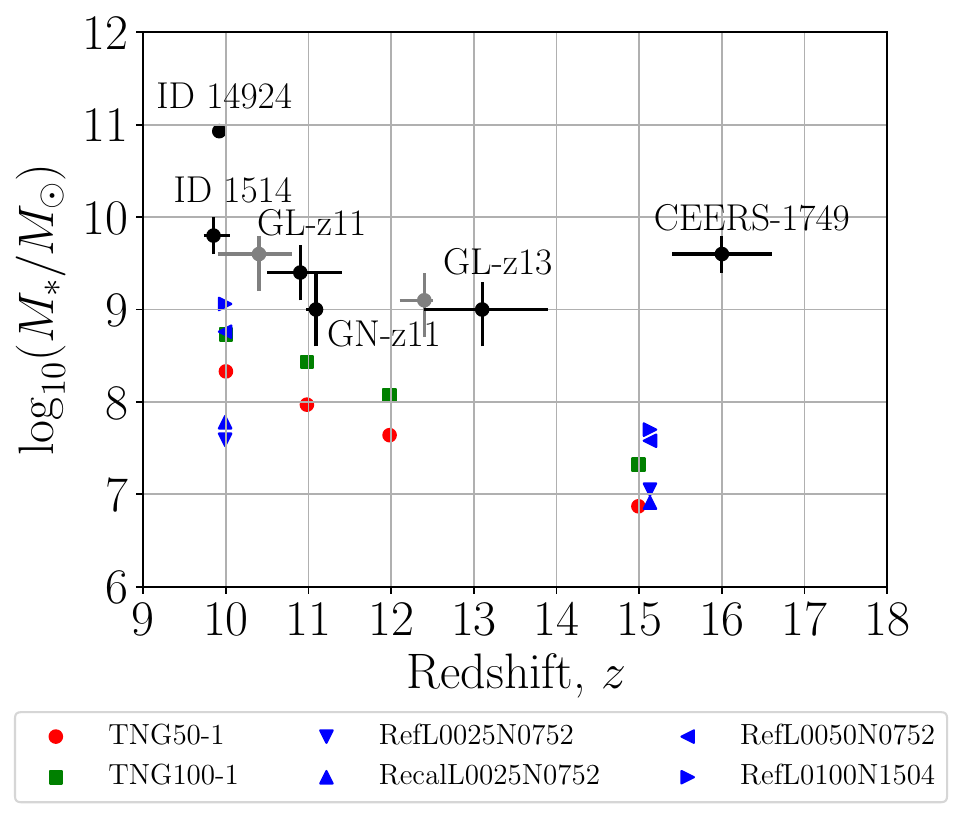}
\caption{The stellar masses of the most massive galaxies already existing at
  redshift, $z=z_{\rm e}$, in the
  $\Lambda$CDM simulations (coloured points). The observed galaxy candidates depicted as
  grey and black errorbars are more than 10 times more massive than
  the most massive galaxies formed in the $\Lambda$CDM simulations
  (coloured points). GN-z11 is a spectroscopically confirmed galaxy
  detected with the Hubble Space Telescope \cite{Oesch+2016}.
  \cite{Naidu+2022a} updated the physical properties of the galaxy
  candidates GL-z11 (now labeled as GL-z10 in \cite{Naidu+2022a}) and
  GL-z13 (now GL-z12) in the accepted version of their paper. Their
  updated redshifts and stellar masses are visualised as grey
  errorbars in the figure. These authors updated these values after Haslbauer et al. (2022
  \cite{Haslbauer+2022b}) got accepted for publication such that
  \cite{Haslbauer+2022b} were not able to include them in their
  analysis. However, the updated values do not change the conclusion
  of \cite{Haslbauer+2022b} in any way as can be seen by the grey
  error bars (GL-z10 and GL-z12). Haslbauer et al. (2022 \cite{Haslbauer+2022b}) also discuss
  if the tension between the observed very early existence of massive
  galaxies can be alleviated if the galaxy-wide stellar IMF (gwIMF) was top
  heavy (thereby leading to smaller real masses), but the calculations suggest that 
  the reduction of the observed masses does not suffice to push them into the SMoC-allowed regime. More calculations with complex star-formation histories are needed to verify this conclusion.
  If the SMoC simulations had been performed with a systematically varying 
  gwIMF then the stronger feedback energy from stellar populations with a 
  top-heavy gwIMF would lead to smaller stellar masses of the model galaxies, 
  thus not solving the problem, since both the observed and the model masses would be reduced.  
  The \cite{Haslbauer+2022b} model stellar populations assume dust-free galaxies 
  (as suggested by \cite{Ferrara+2022} to be the case). Credits: Modified version of
  fig.~2 in \cite{Haslbauer+2022b}.  }
\label{fig:JWST}
\end{figure*}

{\bf (pT3)} A critical aspect constraining any model of structure
formation is {\it how quickly the first galaxies emerge}. The
observations performed with the James Webb Space Telescope (JWST)
indicate that the real Universe appears to have already formed massive
galaxies with stellar masses $\approx 10^{9-11}\,M_\odot$ at redshifts
$10 \simless z_{\rm e} \simless 20$, i.e.  much earlier than possible in the
SMoC (Haslbauer et al. 2022 \cite{Haslbauer+2022b}), 
as shown in  Fig.~\ref{fig:JWST} (see also
\cite{Arrabal+2023} on spectroscopic confirmation).  SMoC-$\Lambda$WDM
models of structure formation come under more stress through such
observations since in these the most-massive galaxies that can form at
a given redshift are less massive than in the SMoC-$\Lambda$CMD models
\cite{DayalGiri2023}.

{\bf (pT4)} Related to this, a possible tension concerning {\it growth
  of structure} from inflation through to the dark-matter-halo mass
function constrained through weak-lensing observations has emerged
(e.g. \cite{Gu+2023}). This problem is related to the ``sigma 8
tension'' and raises questions concerning the empirical normalisation
(based on the abundance of galaxy clusters,
e.g. \cite{BunnWhite1997}) of the power spectrum of perturbations
generated during inflation and how this transfers during linear
structure growth to the power spectrum of matter at $z_{\rm e} \approx 130$
used to seed cosmological structure formation simulations
(e.g. \cite{Vogelsberger+2014}).

{\bf (pT5)} Only about 50~per cent of the baryons that ought to be in the universe according to the SMoC have been discovered (e.g. \cite{LEM_2023}). While this is seen as an important problem for the SMoC with the understanding that the missing baryons are in the filaments and between galaxies awaiting to be discovered, this "missing baryon" problem may simply be a consequence of the KBC void or an even larger-scale inhomogeneous matter distribution (T6, T7 above). 

Reviews of the theoretical approaches to cosmology in view of the
above tensions are available in \cite{Baryshev+1994},
\cite{Buchert+2018, LCorredoira2022} and \cite{LCorredoiraMarmet2022}.
An independent review of the tensions and failures of the SMoC, given astronomical observations, is provided by Peri \& Skara (2022 \cite{PeriSkara2022}),
while Banik \& Zhao (2022 \cite{BanikZhao2022}) provide an assessment of these in
comparison to a Milgromian universe.  \\

\begin{figure*}
    \centering
    \vspace{-25pt}
    \includegraphics[width=0.99\columnwidth]{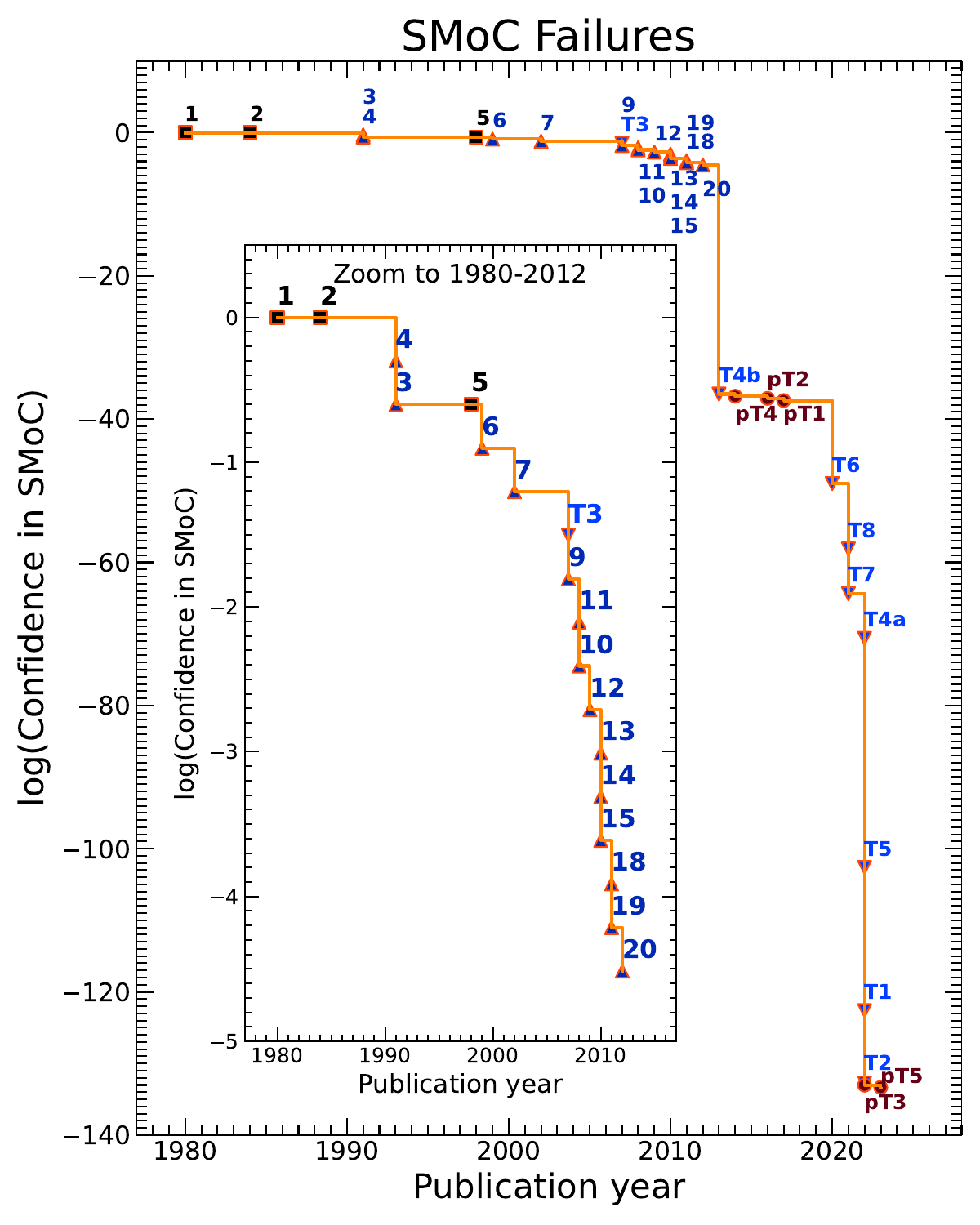}
    \caption{The SMoC-Confidence Graph: the cumulative loss in
      confidence that the Standard Model of Cosmology (SMoC) is a
      valid description of nature. The numbers 1-20 are based on a
      previous review (Kroupa, 2012, \cite{Kroupa2012}), where an
      original form of the current plot appeared. Black squares (1, 2,
      and 5, representing inflation, dark matter, and dark energy,
      respectively) are treated in the SMoC as ``new physics'', so
      they are not assigned a loss of confidence. Upward blue
      triangles indicate failures, still current, already recognized
      in \cite{Kroupa2012}, while downward blue triangles (T1--T8)
      represent newly identified tensions where the loss of confidence
      was computed formally, as presented in Section
      \ref{sec:SMoC_tensions}. From the same section come the possible
      tensions (pT1--pT5), shown with red circles. Wherever the loss
      of confidence was not computed formally, we assign a drop in
      confidence by 50\%. The inset graph zooms into the
      falsifications up to 2012. }\label{fig:confidencegraph}
\end{figure*}

{\it In Summary}: taken at face value, the above compilation
demonstrates there to be no dark matter on the scales of galaxies, and
the real Universe to be significantly more inhomogeneous on scales up
to a few~Gpc than allowed by the SMoC.  The SMoC thus fails on
essentially all scales. This failure can be visualised using the
theory-confidence graph \cite{Kroupa2012}. Based on the
SMoC-confidence graph originally shown in fig.~14 of Kroupa (2012,
\cite{Kroupa2012}), we present an updated version in
Fig.~\ref{fig:confidencegraph}. Fig.~14 from \cite{Kroupa2012} is now
the inset plot in our new figure. The numbers of the ``tensions''
known in 2012 are retained and the reader is directed for a
description of these in \cite{Kroupa2012}. None of these pre-existing
tensions have been resolved as of the current year.  Since black
squares (1: inflation, 2: dark matter, and 5: dark energy) denote
phenomena that are treated as "new physics" in the SMoC, we exclude
them from contributing to the loss of confidence. Some of the older
tensions presented in \cite{Kroupa2012} are here overwritten by more
modern tests, so they are absorbed in the new tensions (T1--T8)
documented in this contribution. Thus we have: tension~8 in
\cite{Kroupa2012} becomes now T4a (i.e. $8\rightarrow{\rm T}4a$), and
likewise $17\rightarrow{\rm T}8$, $16 + 21\rightarrow{\rm T}6$,
$22\rightarrow{\rm T}9$.  For most of the tensions documented in
Sec.~\ref{sec:SMoC_tensions} there are formal confidence results, and
the downward drop corresponds to this amount of loss of confidence
(e.g.: a $5\,\sigma$ tension is displayed here as a drop in confidence
by 6~orders of magnitude). For the ``possible tensions'' (pT numbers)
there are no formal confidence results and thus each is taken to
correspond to a drop in confidence by 50~per cent.
    
{\it The fundamental cosmological principle}: 
Thus, as discussed above, observational evidence suggests that (i) the SMoPP is valid on all observed scales, but (ii) there is no sufficiently large scale where the Universe is homogeneous, i.e, the cosmological principle has not been verified on the scales reached by observation. 

Therefore, all cosmological models based on the cosmological principle of homogeneity and isotropy (i.e., all models based on FLRW geometry) may be flawed. Future efforts to develop reliable cosmological models should consider in earnest the evidence for inhomogeneities on all scales. Given the detrimental performance of the SMoC, in the following Sec.~\ref{sec:alternatives} some alternative models are discussed.

\section{Towards alternative models}
\label{sec:alternatives}

Among the cosmological models which have been proposed over the years, we will be discussing some of those which challenge the three pillars of cosmology (inflation, dark matter, and dark energy) as well as the assumption of homogeneity and isotropy.
Thus, in the following Sec.~\ref{sec:GRbased}--\ref{sec:AMoC} a number of cosmological models are discussed that are relevant for the tensions documented in Sec.~\ref{sec:SMoC_tensions}. Table~\ref{tab:models} gives an overview of these.

\begin{table}[ht]
\begin{centering}
\begin{tabular}{c|ccccc}
Model                                                           & \multicolumn{1}{c}{inflation?} & \multicolumn{1}{c}{DM?} & \multicolumn{1}{c}{dark energy?} & \multicolumn{1}{c}{role of CMB}  &\multicolumn{1}{c}{simulations?}                                                                  \\ \hline
SMoC                                                            & yes                             & yes                               & yes                               & SoLS     &many                                                                                        \\ 
timescape                                                       & yes                               & yes                               & no                                & SoLS     &no                                                                                        \\ 
\begin{tabular}[c]{@{}c@{}}MMoC\\ \small{($R_{\rm h}=ct$)}\end{tabular} & no                              & yes                               & no                                & \begin{tabular}[c]{@{}c@{}}SoLS \\ with dust reprocessing\\ at $z_{\rm e}\approx 16$\end{tabular}  &no\\ 
SIV                                                             & in progress                               & no                                & no                                & in progress        & no                                                                                        \\ 
AeST                                                            & yes                             & no                                & yes                               & SoLS     &no                                                                                        \\ 
\begin{tabular}[c]{@{}c@{}}AMoC\\ \small{($\nu$HDM)}\end{tabular}       & yes                             & no                                & yes                               & SoLS      &a few                                                                                       \\ 
BMoC                                                            & no                              & no                                & no                                &  \begin{tabular}[c]{@{}c@{}}starlight ($z_{\rm e} < z_{\rm *form} > 20$) \\ with dust reprocessing\\ at $z_{\rm e}\simless z_{\rm *form}$ \end{tabular} &starting \\ 
\end{tabular}
\caption{The models discussed here. Each row represents the following cosmological models: discussed in Sec.~\ref{sec:SMoC}: the Standard Model of Cosmology (SMoC), in Sec.~\ref{sec:GRbased}: models based on inhomogeneity (timescape), the Melia $R_h=ct$ model (MMoC), the scale-invariant vacuum model (SIV), the Aether Scalar Tensor model (AeST), in Sec.~\ref{sec:nonN} and~\ref{sec:AMoC}: the Angus model (AMoC, alternatively known as $\nu$HDM) and in Sec.~\ref{sec:conclusions}: the Bohemian model (BMoC). "DM" stands for cold, warm or fuzzy dark matter which is introduced by the respective model to account for the mass deficit on galaxy and cosmological scales. 
"SoLS" is the surface of last scattering, i.e. the photosphere of the Hot Big Bang. $z_{\rm *form}$ is the redshift at which stars begin to form. The column "simulations?" lists if computer simulations of cosmological structure formation are available.}
\label{tab:models}
\end{centering}
\end{table}

\subsection{Based on General Relativity / Newtonian gravitation}
\label{sec:GRbased}

The tensions (T1--T8) between the observations and the SMoC listed in
Sec.~\ref{sec:SMoC} thus comprise pc~to~Gpc scales and at all
redshifts. These tensions imply that more exploration of the SMoC is
called for. For example: \cite{Lahav2023} argue, given that detailed
observations are consistent with the dark-matter plus dark-energy SMoC
(in contradiction to the evidence reviewed above), the increasing data
volumes need advances in machine learning to cast new light on the
cosmological tensions.  \cite{FreeseWinkler2023} propose extensions by
including a dark big bang. Schiavone et al. (2023
\cite{Schiavone+2023}) suggest that a $f(R)$ dark energy model can
lead to an apparent variation of the Hubble constant thus accounting
for the Hubble tension (T6 above). \cite{Ezquiage+2022} suggest that
non-Gaussian exponential tails in inflationary perturbations generated
by primoridal quantum diffusion can account for the observed
large-scale inhomogeneities and the existence of El~Gordo-type galaxy
clusters. All of these suggestions retain dark matter as an essential
pillar of the models and do not address the falsifications of its
presence reviewed above.

The deduction that dark energy drives an accelerated expansion is
dependent on how to time- and spatially-average observational
quantities in an evolving inhomogeneous Universe \citep{Buchert2000,
  Buchert+2023}. The calculations suggest the following: if the
Universe is thought to be homogeneous and isotropic, while in reality
significant inhomogeneities develop over time in a model without dark
energy, then the inferred observational quantities appear as if the
Universe has dark energy and an accelerating expansion (Wiltshire
2007a,b, 2009, 2019, \citep{Wiltshire2007a, Wiltshire2007b,
  Wiltshire2009, Wiltshire2019}).  Wiltshire describes filaments
(walls) and voids with separate homogeneous metrics, and their
statistical average at large scales simplifies to a statistical
homogeneity\footnote{Note that gravity in a homogeneous and isotropic
  metric, such as the FLRW metric, will produce inhomogeneities at
  sufficiently small scales. This is distinct from the inhomogeneity
  and anisotropy intrinsic to the geometry itself, proposed e.g. in
  the Lemaitre-Tolman-Bondi (LTB) metric
  \cite{Lemaitre1933,Tolman1934,Bondi1947,Enqvist2008}.}. An
interesting aspect of the timescape cosmological model is that the
ages of different regions diverge such that in the model universe
voids are older than galaxy clusters. The same oldest stars will thus
appear older in voids than in dense regions, and measuring the global
age of the universe is challenging and can only be achieved
statistically.

This {\it timescape cosmological model} thus suggests dark energy to
be an apparent effect and addresses (but we do not yet know if it also
accounts) for tensions T6--T8. The timescape cosmological model makes
a good case for dark energy to not be real, but it does not account
how the large density contrasts can develop in a self-consistent model
that has the observed CMB as an intial boundary condition.  The
reality of dark energy is also questioned by the analysis by Mohayaee
et al. (2021 \cite{Mohayaee+2021}) of the directionality of the bulk
flow of galaxies and of the measured $H_0$ value and the lack of
convergence to the CMB frame with increasing distance. A
directionally-dependent value of the dark energy density parameter
$\Omega_\Lambda$ based on SN1a data which aligns statistically
significantly with the CMB dipole was previously reported
independently by Javanmardi et al. (2015 \cite{Javanmardi+2015}),
supporting the conclusions reached by \cite{Mohayaee+2021}.

Since the light-horizon ($R_{\rm h}$, the distance travelled by light
over the age of the universe) equals the gravitational radius of the
universe, Melia \& Shevchuck (2012 \cite{MeliaShevchuk2012}) emphasise
that the coincidence is ``{\it disturbing}'' as it can only occur once
in the SMoC (and, by implication, in any model such as the AMoC
insisting on the same expansion history as the SMoC).  They also point
out that ``{\it this equality may actually be upheld for all cosmic
  time $t$ which, however, would not be entirely consistent with
  $\Lambda$CDM, or any other cosmological model we know of.}'' They
therefore argue that the observed Universe expands according to
$R_{\rm h}=c\,t$ such that the gravitational radius and the
light-horizon radius are equal all the time (see also Melia 2012
\cite{Melia2012}). This expansion fits the~SN1a standard candles as
well as the SMoC \citep{Melia2012b}. Neither dark energy nor an
inflationary epoch are needed in this Melia model of cosmology (MMoC,
\cite{Melia2013}), and it therefore may allow the formation of
galaxies at higher redshift than the SMoC \citep{Melia2023}, avoiding
possible tension pT3 (Sec.~\ref{sec:SMoC_tensions}).  The MMoC thus
appears to provide an interesting alternative to the standard
expansion history of the SMoC (that has an initial very brief
inflationary growth, then coasting expansion that slows due to
eigengravity and then accelerates again due to dark energy) by
avoiding fundamental tensions~FT1 and~FT2
(Sec.~\ref{sec:SMoC_tensions}).  Within the MMoC paradigm, interesting
insights as to the origin of the observed CMB are emerging in terms of
the primordial power-spectrum \citep{Melia2019}. While in the MMoC the
observed CMB is interpreted to form at $z_{\rm e}\approx 1100$ as in
the SMoC as the photosphere of a Hot Big Bang (but without inflation),
Melia (2022 \cite{Melia2022}) argues that the observed CMB
anisotropies result from a dust screen formed from the first star
formation near $z_{\rm e}\approx 16$, therewith possibly accounting
for~pT2.

While the work on the timescape cosmological model and the MMoC needs
to be kept in mind in the following, these models require dark matter
to dominate the matter content by being based on Newtonian
gravitation.  That is, they retain~FT3 and~FT4 and thus suffer from
tensions T1-T5 that are a result of needing dark matter to dominate
the matter content of the model universe.
Given the track-record of verified predictions of the MOND paradigm 
as emphasised in the award-winning book of Merritt (2020 \cite{Merritt2020}), 
see also Appendix~\ref{sec:MOND} below), it may appear useful to study
models of structure formation being based on this general framework. 

In this context, Maeder (2017a,b,c \cite{Maeder2017a, Maeder2017b, Maeder2017c}, 
see also Gueorguiev \& Maeder (2022  \cite{GueorguievMaeder2022}) study cosmological models 
based on the hypothesis that macroscopic empty space 
is scale-invariant, a property related to MOND (Milgrom 2009 \cite{Milgrom2009}). 
This Scale-Invariant Vacuum (SIV) theory is based on Weyl's Integrable 
Geometry, furnished with a gauge scalar field. The main difference between 
Milgromian dynamics and the SIV theory is that Milgromian dynamics is equivalent to 
a global scale invariance of space and time, where the scale factor 
is a constant \cite{Milgrom2009}, while  in the SIV theory the scale factor 
is a function of time \cite{MaederGueorguiev2020}. 
Maeder (2023, \cite{Maeder2023}) points out that the SIV theory tends to MOND in the
weak field limit.  The SIV model of cosmology keeps the physical
properties of General Relativity in a model universe void of matter, being based on 
scale invariance, homogeneity and isotropy of empty space. 
The scale invariance, given by the absence of matter such
that a scale cannot be defined, is broken in the presence of matter
and a characteristic acceleration scale emerges that is similar to
Milgrom's constant $a_0$ as a consequence of the equilibrium of the
SIV and Newtonian dynamical accelerations. This approach leads to the
deep-MOND limit being an approximation of the SIV theory at low
densities and for systems with time-scales shorter than a few~Myr.
Cosmological models based on SIV have an accelerated expansion but no
dark energy, while flat rotation
curves of galaxies are given by the space-time scale invariance,
similarly as in MOND, thus avoiding the need for dark matter (and
therewith solving T1). According to \cite{BanikKroupa2020b} SIV is
challenged by Solar System data, and it appears difficult for the
theory to account for the asymmetric tidal tails of star clusters
(T2). Also, the accelerated expansion inherent to the model 
raises the question of energy conservation since space is associated with quantum vacuum energy (FT2). 
The requirement for homogeneity and isotropy of the SIV theory
appears to be in tension with T6--T8. Inflation in SIV
is being studied (Maeder \& Gueorguiev 2021 \cite{MaederGueorguiev2021}). 
The role of the CMB has not yet been explored in this theory (Maeder, private communication), but 
the growth of density fluctuations has been studied
with favourable results \cite{MaederGueorguiev2019}, 
the analytical solution for the early Universe being documented by Maeder (2019 \cite{Maeder2019}).  
More detailed tests are needed, and
these would be best approached with simulations of structure formation. 
This requires equations of motions to be available to integrate
particles through time \cite{Maeder2017a}.

\subsection{Based on non-Newtonian models}
\label{sec:nonN}

The level of the tensions on well observed galaxies suggest, when
taken at face value, a need to explore models of structure
formation that differ fundamentally from the SMoC, the timescape
cosmological model and the MMoC, in order to open up or to reject
possible new directions of investigation.  

In order to avoid the tensions emanating from dynamical friction on
dark matter particles in a Newtonian universe, a different theory of
gravitation might be tried. Various alternatives to Newtonian dynamics
have been suggested (e.g. emergent gravitation related to the entropy
of space, \cite{Verlinde2017}; gravitational dipoles,
\cite{Hajdukovic2019}; tensor-scalar-vector modified gravity,
\cite{MoffatToth2021}; scale-invariant dynamics,
\cite{Maeder2023} ). But these formulations do not readily allow
the equations of motion of particles to be written down for numerical
integration thus inhibiting structure formation simulations down to
the scale of galaxies, and/or they have been found to be in tension
with observational data (respectively: \cite{Lelli+2017b,
  BanikKroupa2020a, BanikKroupa2020b, Haghi+2019}).

A hint which models might
be feasible to study in a first step towards this goal is the inferred 
absence of dark matter particles (T1) together with the documented
success of MOND \citep{Milgrom1983a,
  Milgrom1983b, Milgrom1983c, BekensteinMilgrom1984} on the scales of
galaxies \citep{FamaeyMcGaugh2012, Kroupa+2012, BanikZhao2022} as well
as its promise on cosmological scales (\cite{Sanders1998} and
\cite{McGaugh1999}).  While Milgromian dynamics is briefly
introduced in Appendix~\ref{sec:MOND}, here the available results are
summarised. These results motivate us to explore cosmological models based on
Milgromian dynamics. 

Briefly, by postulating that Newtonian gravitation (represented by the
$p=2$ p-Laplace operator in the Poisson equation) shifts to the $p=3$
p-Laplace operator when the gradient of the potential falls below
Milgrom's constant
$a_0=1.21^{+0.27}_{-0.27} \times 10^{-13}\,
$km/s$^2=3.90^{+0.86}_{-0.86}$pc/Myr$^2$ \citep{Begeman+1991}, one
obtains a theory of gravitational dynamics which is, as far as all
tests so far conducted, consistent with the data.  By being able to
calculate the local gradient of the gravitational potential from the
matter density field, Milgromian dynamics thus allows the equations of
motions of particles to be written down and integrated in a
computer. This is an essential aspect of any new model of cosmology we
might want to study.  Assuming the equations of motion of particles to
be based on Milgromain dynamics has immediate implications for the
tensions catalogued in Sec.~\ref{sec:SMoC_tensions}:

(T1) Dynamical dissipation is not active because dark matter particles
are not needed to boost gravitation, and the large number of perturbed
dwarf galaxies in galaxy clusters at any time are natural in
Milgromian dynamics (Fig.~\ref{fig:Fornaxdwarfs}, as shown by Asencio
et al. 2022 \citep{Asencio+2022}). UDGs are natural in MOND
\cite{Haslbauer+2019a}, but this needs to be verified with
cosmological simulations.

(T2) On the pc-scale, the evidence that stars evaporating from open star
clusters preferentially populate their leading tidal tail in Milgromian
dynamics as observed supports the necessity to switch away from the
Newtonian equations of motion (Kroupa et al. 2022 \cite{Kroupa+2022}).

(T3) Recent hydrodynamical simulations with star formation of
collapsing post-Big-Bang gas clouds have shown rotating disk galaxies
with radial exponential density profiles with the observed physical
sizes emanating naturally (Wittenburg et al. 2020
\cite{Wittenburg+2020}, Nagesh et al. 2023 \cite{Nagesh+2023}).  The
observed distribution of baryons allows an exact prediction of the
rotation curves of galaxies in MOND (e.g. Lelli et al. 2017
\cite{Lelli+2017a}).

(T4, T5) Computations of interacting galaxies have shown galaxy
mergers to be much rarer in Milgromian dynamics since dynamical
dissipation on extended dark matter halos is non existent
\cite{Renaud+2016}. Galaxies can thus orbit about each other multiple
times without merging. This implies that galactic disks remain (even
if warped).  If a galaxy forms from a typically rotating gas cloud
(see T3 just above) then the vast preponderance of disk galaxies would
follow naturally.  The disks/planes of satellite galaxies are
naturally produced in galaxy--galaxy encounters as populations of
co-orbiting tidal dwarf galaxies (Banik et al. 2018 \cite{Banik+2018},
Bilek et al. 2018, 2021 \cite{Bilek+2018, Bilek+2021}, Banik et
al. 2022 \cite{Banik+2022}, see also
\cite{OkazakiTaniguchi2000}). This also implies galactic bars to
remain fast \citep{Roshan+2021}. The dual dwarf galaxy theorem is
automatically obeyed and is consistent with the observed dwarf galaxy
population because all dwarf galaxies (tidal tail and primordial)
follow the same dynamical laws and become indistinguishable apart from
youth and metallicity differences \cite{Kroupa2012, Haslbauer+2019b}.
Polar ring galaxies have shown rotation velocities of their extended
polar rings that are systematically larger than the rotation speed of
their host which arises naturally in MOND (Lueghausen et al. 2013
\cite{Lueghausen+2013}).

(T5) The downsizing time-scales of the rapid formation of elliptical
galaxies arise naturally from initial non-rotating gas clouds (Eappen
et al. 2022 \cite{Eappen+2022}). Sub-grid physics describing star
formation and gas heating and cooling are not critical for these
outcomes (Wittenburg 2020 \cite{Wittenburg+2020}, Eappen et al. 2022
\cite{Eappen+2022}, Nagesh et al. 2023 \cite{Nagesh+2023}) such that
cosmological structure formation simulations in Milgromian dynamics of
a largely baryonic universe with galaxy-scale resolution are expected
to lead to a population of galaxies that should have the correct
properties. The rapid formation of spheroids (bulges, elliptical
galaxies) produces the exact environment in which SMBHs form naturally
and fast (Kroupa et al. 2020 \cite{Kroupa+2020}), while the rarity of
mergers leaves these at the centres of their hosting galaxies as
observed (Lena et al. 2014 \cite{Lena+2014}).

(T6, T7) The existence of, the size and depth of KBC-like voids arise
naturally in a Milgromian universe (Fig.~\ref{fig:HT1} here, Haslbauer
et al. 2020, \cite{Haslbauer+2020}). Concerning the Hubble Tension and
the general problem of how quickly the overall Universe expands and
thus how old it is, the oldest observed stars provide a
cosmology-model-independent measure of the local age of the
Universe. Their ages are reported to be consistent with the global
value, $H_0^{\rm global}$ (see Sec.~\ref{sec:conclusions}, with a
tendency towards an older local Universe). The stellar ages thus
indicate that the Planck cosmology may be closer to the
background-level expansion.  This means that the Hubble Tension is
most likely due to the galaxies having developed peculiar velocities
as they fall towards the matter overdensities surrounding the KBC void
(Haslbauer et al. 2020 \citep{Haslbauer+2020}). Solutions involving
more complex dark energy models, such as early dark energy, that would
affect the age of the universe and thus the ages of the oldest stars,
are therefore not necessary. Systematic large-scale matter motions of
nearly~$1000\,$km/s over 500~Mpc as reported by Migkas et al. (2021
\cite{Migkas+2021}) appear to be naturally obtained in MOND-based
cosmological models of structure formation (Candlish 2016
\cite{Candlish2016}).

(T8) The existence of the El Gordo and Bullet clusters with their
respective masses and redshifts appears to be well accounted for in a
Milgromian universe (Katz et al. 2013 \cite{Katz+2013}, Asencio et al. 2021 \cite{Asencio+2021}). 
A minor tension has emerged
in MOND with the gravitating masses of galaxy clusters being larger by
about a factor of two than the observed baryonic masses
(Sanders 1994 \cite{Sanders1994}). However, this tension may be due to the previous
analyses lacking pressure corrections and/or having used inadequate
profiles for the baryonic matter content
(Lopez-Corredoira et al. 2022 \cite{Lopez-Corredoira+2022}).

But some tension between MOND and the data have emerged as follows. MOND Tension~1 {\bf (MT1)}: Ultra-faint dwarf satellite galaxies (UFDs)
have half-mass radii between
$20\simless r_{\rm h}/{\rm pc} \simless 500$ and masses in stars
between about $10^{2.5}$ to $10^4\,M_\odot$ (e.g. \cite{Wolf+2010, Simon2019}). Safarzadeh \& Loeb (2021
\cite{SafLoeb2021}) show that the observed line-of-sight velocity
dispersion in these objects is approximately constant between about
4~to $10\,$km/s while it should be decreasing with the mass in stars
to values below $1\,$km/s. The analytical estimates of the theoretical
velocity dispersion taking into account the external field of the
Galaxy may not be correct though, and self-consistent simulations of
the evolution of UFDs are needed to address this tension with
rigour. This is needed since simulations of such objects done in
Newtonian gravitation have uncovered hitherto unknown solutions of
quasi-stable remnants of satellite galaxies that are not in virial
equilibrium and appear to be dominated by dark matter but contain no
dark matter (Kroupa 1997 \cite{Kroupa1997}, see also \cite{Casas+2012}). 
Such work needs to be redone in
Milgromian dynamics.

{\bf (MT2)}: Based on hydrodynamical simulations of isolated disk
galaxies that are based on initial conditions given by present-day
observed galaxies and that allow for star formation to develop in the
simulations, Nagesh et al. (2023 \cite{Nagesh+2023}) calculate the pattern speeds of the
bars that form in these models. As for real galaxies, they find the
bar length to be larger and the bar to be weaker for the more massive
models, but bar pattern speeds, while being constant in time, to be
slower than inferred for the real galaxies and in particle-simulations
only. Formally the models disagree with the observed bar pattern
speeds at more than $5\,\sigma$ confidence, but a rigorous comparison
needs to be done with respect to galaxies formed in cosmological
structure formation simulations, as was done above for the SMoC case
(T1).

{\bf (MT3)}: Tension between MOND and data is also evident for the
globular cluster NGC~2419 \cite{Ibata+2011}, but detailed stellar
dynamical modelling \cite{Sanders+2012} taking it's possible birth
conditions into account may alleviate these \cite{Kroupa+2022}.

{\bf (MT4}): Also, some tension is evident in the Solar System with
the external field from the Galaxy leading to the secular precession
of the perihelion of Saturn in combination with that of Earth not
favouring the usual interpolating function needed for modelling galaxies
between the weak-field MOND and strong-field Newtonian regimes
\cite{BlanchetNovak2011}, but other interpolating functions are
consistent with the Solar system and galactic data
\cite{Hees+2016}.

{\bf (MT5}): Furthermore, the GAIA wide-binary-star data (binaries
with separations larger than a few thousand astronomical units) have,
as applied so far, been suggesting non-conformity with Newtonian
dynamics, and appear to be in agreement with MOND (\cite{Scarpa+2017,
  Hernandez+2022, PittordisSutherland2023, Chae2023}, Banik et al.,
submitted). This test sensitively dependents on the multiplicity
properties of the low-mass stars being used (Clarke 2020
\cite{Clarke2020}) as well as on stellar astrophysics and evolution,
such as the stellar mass-luminosity relation and the possible pre-main
sequence nature of the stars. A pre-main sequence star will be
calculated to have a larger mass if it is wrongly thought to be a main
sequence star, and so this bias hides the effectively stronger
Milgromian gravitation.  Further work is needed before final
conclusions can be reached, but the most recent analysis of the
Gaia~DR3 wide-binary data by Chae (2023 \cite{Chae2023}) strongly
supports Milgromian gravitation.

{\bf (MT6}): The Large Magellanic Cloud (LMC) is currently passing the
MW at near-pericentre of its orbit at a distance of about 50~kpc from
the MW. While the mass in stars and gas of the LMC
($\approx 4\times 10^9\,M_\odot \approx 5\,$per cent of the MW
baryonic mass) is consistent with the baryonic Tully-Fisher relation,
given the observed rotation curve, its gravitating mass as deduced
from the reflex motion of the MW suggests a larger gravitating mass
(10--20~per cent of the MW, \cite{Vasiliev2023}). Does this suggest
that some dark matter needs to be present to boost the gravitating
mass of the LMC, therewith breaking the success of MOND? This
deduction would be premature because the expected reflex motion of the
MW needs to be calculated in MOND taking into account that the fully
self-gravitating MW disk is breathing and oscillating in 3D which
might feign some of the reflex motion. The same is true for the
stellar streams that also constrain the gravitating mass of the LMC
\cite{Vasiliev2023}.

{\it In Summary}, from pc~to~Gpc~scales, nature appears to comply to
Milgromian dynamics rather than Newtonian dynamics plus dark matter, a
deduction also reached by Banik \& Zhao (2022 \cite{BanikZhao2022})
based on an independent assessment. In his award-winning book, Merritt
(2020 \cite{Merritt2020}) applies the formalism of the philosophy of
science to address the merit of the MONDian research programme in
comparison to that of the SMoC programme, finding that only MOND has
repeatedly predicted observational facts in advance of their
discovery. Based on the criteria as applied in the philosophy of
science, MOND is therewith the more viable and a progressive research
programme.

Adopting Milgromian dynamics avoids most of the tensions of
Sec.~\ref{sec:SMoC_tensions}. A few tensions have emerged as noted
here, but these need rigorous further analysis in order to ascertain
if they might lead to a falsification with at least $5\,\sigma$
confidence of MOND in general, or of MOND in terms of its
interpretation as a gravitational theory or as a theory of modified
inertia. Also, at the moment it remains unclear whether the correct
fraction of elliptical versus disk galaxies will truly emanate from
structure formation in Milgromian dynamics. It is also unclear if a
Milgromian-based cosmological model will account for the observed
Gpc-scale density inhomogeneities and bulk matter flows.  Self-consistent cosmological
simulations are required to directly address this question as well as
whether the distribution of galaxies in phase space and in galaxy
clusters comes out comparable to that in the observed Universe.

The construction of a fully self-consistent cosmological model based
on one set of relativistic field equations that quantify how matter
curves space dynamically would be needed. This underlies the
interpretation of gravitation as a strictly geometric
effect.\footnote{\label{ftn:gravitation} Gravitation is not well
  understood and according to another interpretation it may be an
  effect emerging from spatial entropy differences
  \citep{Verlinde2011, Verlinde2017}, or it might arise as a
  consequence of the wave-nature of matter \citep{Stadtler+2021,
    Zhang+2022}.}  The recently developed new relativistic theory of
modified Newtonian dynamics (the Aether Scalar Tensor -- AeST) model
by Skordis \& Zlosnik (2021 \cite{SkordisZlosnik2021}) proposes such a formulation and accounts
for the standard CMB, the standard matter power-spectrum in the linear
regime and that gravitational waves propagate with the speed of electromagnetic waves, $c$. It reproduces
both the putative successes of the SMoC on cosmological scales and
those of MOND on galactic scales.  The AeST model requires an initial
inflationary epoch, all matter being created during the Hot Big Bang, and dark energy to account for
the observed expansion rate, as deduced from the observed CMB being
interpreted as the surface of last scattering and from the SN1a data. It has a very
similar expansion history as the SMoC because the AeST model satisfies the
usual Friedmann equations. It is therewith subject to fundamental
tensions~FT1 and~FT2.  The AeST model fails to account for the observed
inhomogeneities as it aims to reproduce the SMoC rather than
observations (tensions T6--T8, \cite{BanikZhao2022}), and has also
been found to encounter problems with weak lensing by galaxies (Mistele 2023 
\cite{Mistele+2023}).  The AeST model requires more investigation, but
here it is not discussed further as at present we do not have a
computer programme that allows the equations of motion of baryons to
be integrated self-consistently in view of the additional fields this
model requires.

Given all the above, here a pragmatic and computationally-accessible approach is
taken by combining theoretical knowledge that may be viewed to be
reliable to provide first steps towards a holistic understanding of
cosmological physics. We begin with the Angus model of cosmology.

\section{\texorpdfstring{The Angus model of cosmology (AMoC) -- the $\nu$HDM model}{The Angus model of cosmology (AMoC) -- the $\nu$HDM model}}
\label{sec:AMoC}

The AMoC constitutes a first conservative and thus careful step towards a Milgromian cosmological model by
retaining postulates A1--A5 and auxiliary hypotheses~AH1 and~AH3. The AMoC removes AH2, i.e. cold or warm dark matter particles, from the computer experiments. In the AMoC the generalised Poisson equation based on
the p-Laplacian which connects the baryonic overdensities with the
Milgromian gravitational potential  is applied rather than the standard Poisson equation (Appendix~\ref{sec:MOND}).  While
postulate~A3 (Einsteinian gravitation is valid globally) seems in conflict with this ansatz, the argument can be made
that Milgromian dynamics emerges despite~A3 due to the quantum vacuum
affecting the motions of matter particles when space-curvature falls
below the acceleration $a_0$ \cite{Milgrom1999}.

In order to match closely the observed CMB and to ensure reproduction of the same expansion history as the SMoC, the AMoC (Angus 2009 
\cite{Angus2009})\footnote{The AMoC is often referred to as the
  ``$\nu$HDM'' model, e.g. Haslbauer et al. (2020 \cite{Haslbauer+2020}).} 
is based on the assumption
that the relic mass density of 11 eV sterile neutrinos\footnote{Note that the possible existence of such sterile
  neutrinos as a hot dissipation-less matter component is motivated by
  neutrino physics rather than only by evidence from cosmological
  observations and galactic dynamics.} 
matches that of CDM in the
standard $\Lambda$CDM model, and that dark energy contributes the remaining energy-density. 
In this model, the sterile neutrinos do not play a dynamical role in galaxies and 
Local-Group-type structures and thus do not lead to Chandrasekhar dynamical friction 
between galaxies, because they behave as hot dark matter and thus only contribute to the masses of galaxy clusters. 
Angus \& Diaferio (2011 \cite{AngusDiaferio2011}) show that further refinements to the 
cosmological parameters are possible (see their fig.~1). Their result suggests 
that finding an optimal fit to the observed CMB in $\nu$HDM would slightly increase 
the inferred $H_0^{\rm global}$, therewith partially lessening the Hubble Tension.
The AMoC is therefore defined by the same six parameters as the SMoC (Sec.~\ref{sec:SMoC})
in order to describe the current state of the model universe in terms
of its expansion rate, mass components, initial density fluctuations
and curvature, with the only difference that $\Omega_c$ is replaced by
the sterile neutrino mass density parameter $\Omega_\nu \approx 0.26$
(see also Katz et al. 2013 \cite{Katz+2013} and 
Wittenburg et al. 2023 \cite{Wittenburg+2023}).  
The inflation-produced CMB-reproducing matter power-spectrum of the AMoC
is very similar to that of the SMoC. Linear structure growth however
enhances the power on small scales in the SMoC due to the presence of
cold or warm dark matter particles that allow gravitational clumping
of these components on small scales already prior to $z_{\rm e}=1100$.
However, by $z_{\rm e}\approx 200$, the matter power spectrum of the AMoC
falls off by orders of magnitude on scales smaller than about 30~Mpc
in comparison to that of the SMoC (fig.~2 in Angus \& Diaferio 2011 \cite{AngusDiaferio2011})
because the sterile neutrinos, being a hot dissipation-less
matter-dominating component together with the baryonic plasma and
radiation component hinder early gravitational clumping on small
scales.  In an AMoC simulation, neither galaxies nor other structures
form in a simulation box smaller than 30~Mpc.  The simulation boxes
therefore need to be larger than 100~Mpc to capture structure
formation since larger structures develop and collapse prior to
smaller ones within them.  The available computations do not yet reach
a resolution sufficient for individual galaxies (Wittenburg et
al. 2023 \cite{Wittenburg+2023}).

Assuming Milgrom's constant $a_0$ is time invariant and noting its
value to be $a_0\approx c\,H_0 / \left( 2\,\pi\right)$ implies that
the universe as a whole transitions into the Milgromian regime at $z_{\rm e}\approx  3$  (McGaugh 1999 \cite{McGaugh1999}), although density contrasts
can develop much earlier as a consequence of the stronger effective
gravitational field in MOND (Sanders 1998 \cite{Sanders1998}). It is safe to assume
structure growth to remain in the Newtonian regime down to 
$z_{\rm e}=200$
when MOND-cosmological simulations are initialised \citep{Nusser2002,
  Katz+2013}.

Pioneering studies of cosmological structure formation in Milgromian
gravitation had already suggested larger and more massive structures
to assemble more quickly than in the SMoC (Nusser 2002, \cite{Nusser2002}). Within
these, Sanders (1998 \cite{Sanders1998}, 2008 \cite{Sanders2008}) argues that
galaxies form earlier and more rapidly than in the SMoC, with voids being much larger and
emptier than in the SMoC. Some of this has been verified with actual
cosmological simulations within the AMoC using dissipation-less
particles to trace the baryonic matter by Angus et al. (2011 \cite{Angus+2011}) and Katz et al. (2013
\cite{Katz+2013}). These dissipation-less particle simulations indicate
however that the present-day mass function of galaxy clusters might be
too flat, extending to too large masses than observed (fig.~5 in 
Angus \& Diaferio 2011 \cite{AngusDiaferio2011}; fig.~1 in Katz et al. 2013 
\cite{Katz+2013}).  As shown by Haslbauer et al. (2020, \cite{Haslbauer+2020}),
this may be
alleviated though by noticing that the Local Group is situated in the
about 0.6~Gpc KBC void such that the amplitude
of the matter fluctuations within it are muted. This may also alleviate the sigma~8~tension of the SMoC (pT4).

\begin{figure*}
    \centering
    \vspace{-18.523pt}
    \includegraphics[width=0.65\columnwidth]{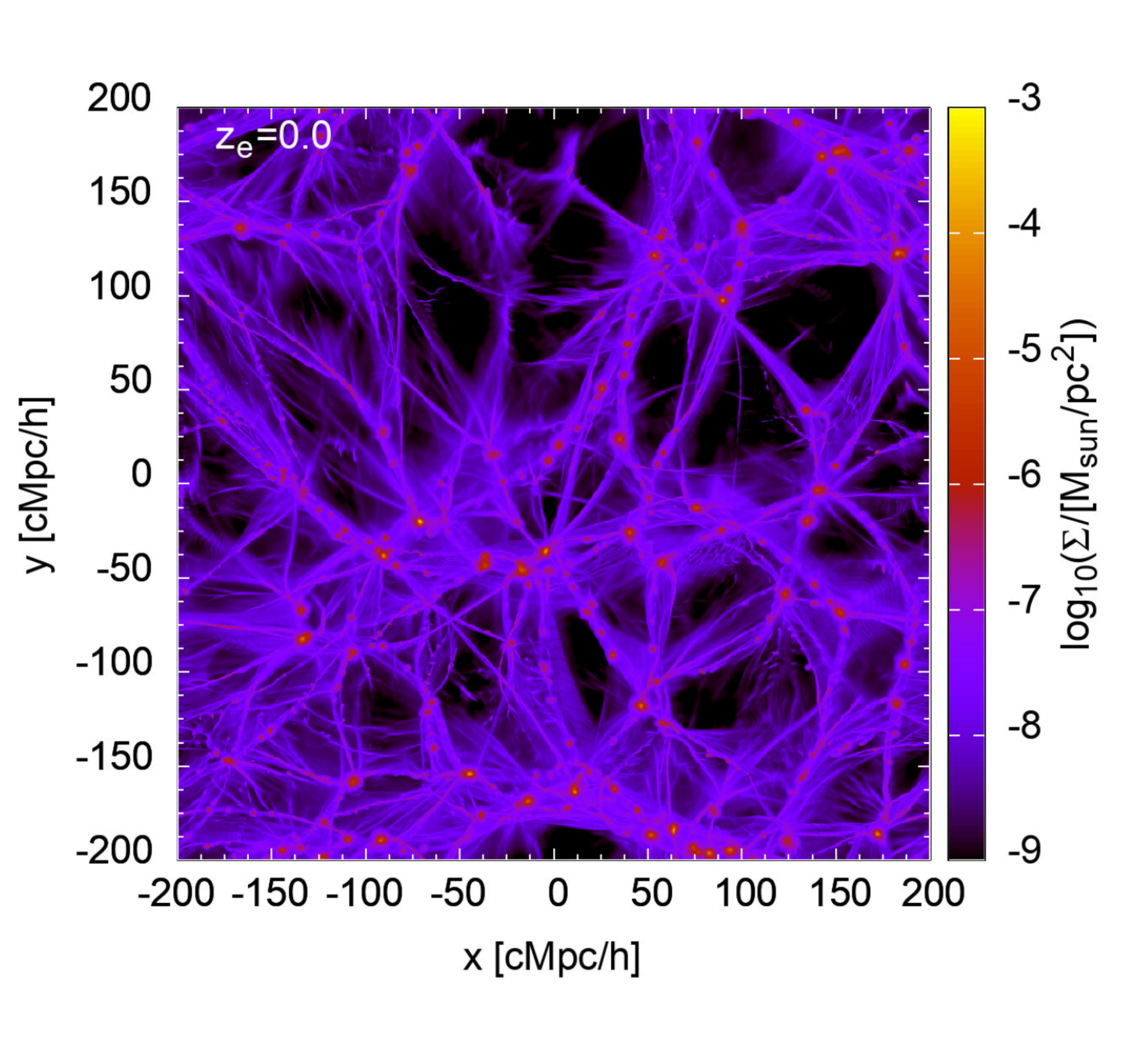}\vspace{-5pt}
    \includegraphics[width=0.65\columnwidth]{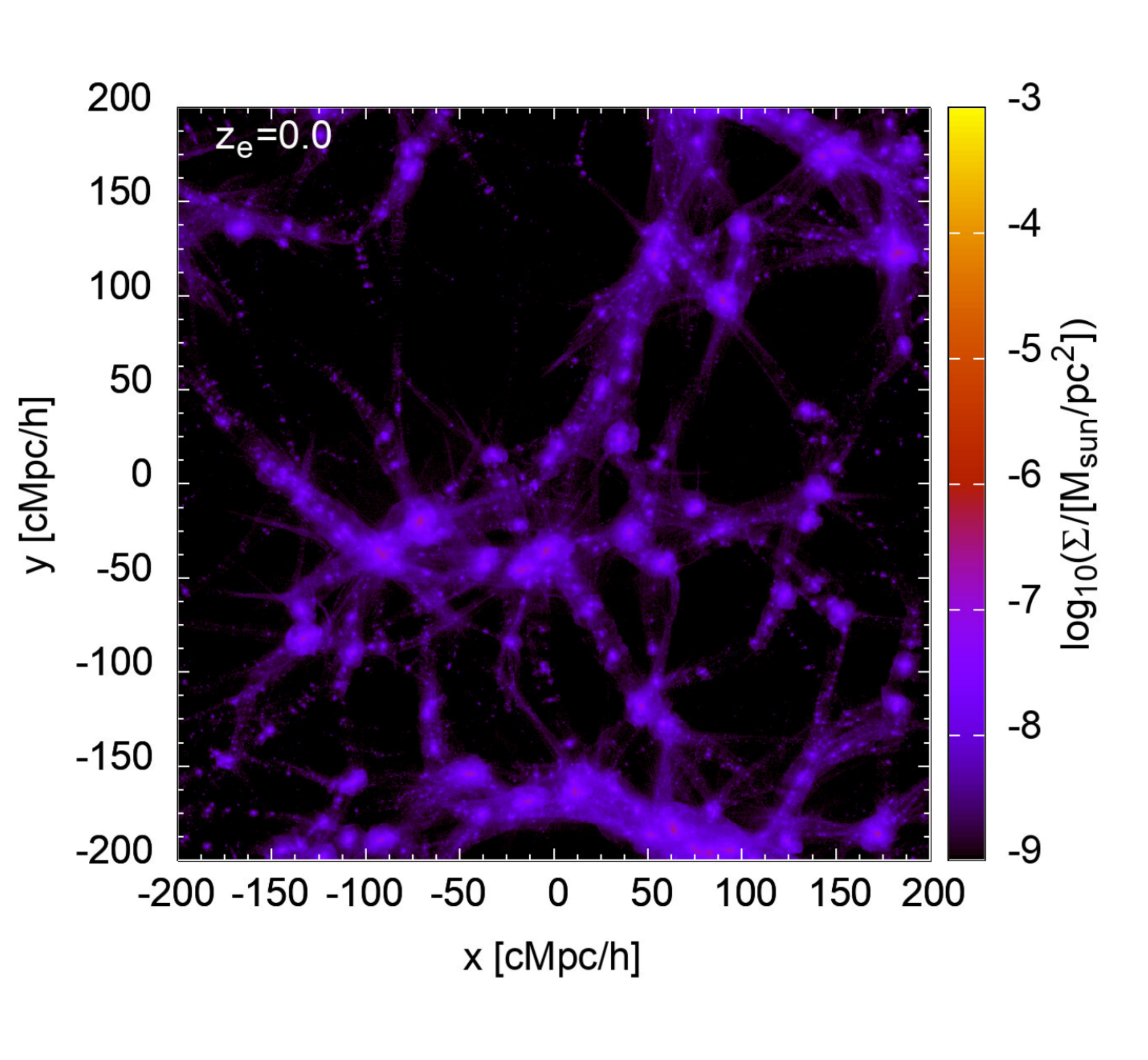}
    \caption{Hydrodynamical simulations of cosmological structure formation in the AMoC (i.e. $\nu$HDM model, $z_{\rm e}=0$ snapshots of a $400\,$co-moving Mpc box). The simulations were performed by Nils Wittenburg in Bonn with the PoR code (Appendix~\ref{sec:MOND}). {\it Upper panel}: the mass density of baryonic matter projected onto the xy plane. The density shown here is the mass-weighted average for the corresponding pixel. {\it Bottom panel}: Similar to the top panel, but for the sterile neutrinos.
    The dense clumps in both panels are galaxy clusters; galaxies are not resolved in these simulations. These simulations show that the mass-fraction of sterile neutrinos to baryons increases towards more massive galaxy clusters, being insignificant on the scale of galaxies (such that only MOND matters for their dynamics) while being consistent with the observationally-deduced fraction \cite{PointecouteauSilk2005} of about 100~per cent for massive clusters. Movies of the simulations are available in Wittenburg et al. (2023 \cite{Wittenburg+2023}). With permission from Wittenburg et al. (2023 \cite{Wittenburg+2023} their fig.~3).}
    \label{fig:Wittsim}
\end{figure*}

The AMoC thus avoids tension (Sec.~\ref{sec:SMoC}) T1, accounts for T2
and appears to alleviate tensions T3--T8 (with its role on T4 being
unclear), although it may overproduce massive galaxy clusters (T8). In
an attempt to further address this problem and also the problem of
galaxy formation in view of tension~T5, modern (first-ever)
hydrodynamical simulations of the AMoC have been performed in Bonn
using the PoR code (Appendix~\ref{sec:MOND}) and are being analysed in
Bonn, Prague and St. Andrews (Wittenburg et al. 2023
\cite{Wittenburg+2023}). These simulations do indicate that
possible tension pT3 (JWST observations of very early galaxy
formation) may worsen in the AMoC, since gravitationally bound
structures are detected only at $z_{\rm e}\simless
4$. Fig.~\ref{fig:Wittsim} demonstrates an example of a $z_{\rm e}=0$
snapshot for the baryonic and sterile neutrino density distribution
obtained with these simulations.

\section{Conclusions}
\label{sec:conclusions}

This contribution is an update of previous 2012--2015 assessments of the SMoC (\cite{Kroupa+2010, Kroupa2012, Kroupa2015}). It transpires that none of the problems previously identified have been
solved, and many new problems arise on ever larger scales (T1--T8, pT1-pT5). 
This is visualised graphically by
Fig.~\ref{fig:confidencegraph}. The SMoC appears to be under
significant stress on all scales and all redshifts.  
There appears little alternative but
to face up to the possibility that 
the SMoC may need to be replaced by
a different model. 
This conclusion, drawn from a multitude of
independent tests of the SMoC against data on all scales, conforms
with the SMoC having the unphysical property that it is energy
non-conserving (FT2, Sec.~\ref{sec:SMoC_tensions}).  But can an
alternative model be constructed?

An important constraint is that any viable  model ought to obey the {\it fundamental cosmological principle} (Sec.~\ref{sec:SMoC_tensions}) by 
allowing for
significant structure growth on all observed scales, i.e. from~kpc to~Gpc
scales, and probably also on the scale of the entire Universe (T7, pT2
in Sec.~\ref{sec:SMoC_tensions}).  This condition discourages most cosmological
models that have been suggested to date, as most have been
constructed on the premise that the model be homogeneous and isotropic
on large ($>100\,$Mpc) scales to be in agreement with the wrongly-surmised
success of the SMoC on these scales. 
Table~\ref{tab:models} lists the models discussed here.

The perhaps most important clue as to how to construct a new model is
the negative tests for the existence of dark matter particles with the
significance of the independent tests amounting to well over the
$5\,\sigma$ rejection threshold of the hypothesis that dark matter
exists (T1, Sec.~\ref{sec:SMoC_tensions}).  Since there is also a lack of motivation
from the standard model of particle physics for the existence of dark
matter particles, it might appear useful to study models without dark
matter.

Given the documented success of Milgromian dynamics (with some
tensions existing though: MT1--MT6, Sec.~\ref{sec:nonN}), one way
forward is to consider constructing cosmological models based on
Milgromian equations of motion rather than Newtonian ones
\cite{Kroupa+2012}.  Milgromian dynamics is, to our knowledge, the only known
other-than-Newtonian formulation of dynamics which is consistent with
Solar System data and also explicitly formulates the
equations of motion to be used to integrate a self-consistent system
through time. Therefore a significant research effort (in Bonn, Prague and
St.~Andrews) is currently concentrating on exploring the $\nu$HDM
model (i.e. the AMoC, Sec.~\ref{sec:AMoC}). This model is traditional
in terms of minimalistic departures from the current SMoC by retaining
the fundamental postulates A1--A5, and also the auxiliary hypothesis
of inflation (AH1) and dark energy (AH3). The model improves the
situation significantly as it avoids the problems associated with star
clusters and galaxies and generates large voids with corresponding
bulk matter flows (thus alleviating T1--T8). But further tests using
simulations need to be made to affirm if this is truly the case. The
possible tensions (pT1--pT5) remain though, and the results available
from the simulations done so far suggest that the AMoC fails on pT3
(model galaxies form by a redshift of $z_{\rm e}\approx 4$ and thus
far too late), and in that the mass function of galaxy clusters may
extend to too large masses and be too top-heavy compared to the
observationally constrained one (Wittenburg et al. 2023 \cite{Wittenburg+2023}).  If each of
the tensions associated with the AMoC correspond to a loss of
confidence by 50~per cent (we do not yet have rigorous tests), then
the MT1--MT6 tensions plus the probable failures with the mass function
of galaxy clusters and the too-late formation of structures amount to
a remaining confidence of 0.4~per cent that the AMoC is a correct
description of the Universe.  Furthermore, the AMoC retains the
problems of the SMoC in terms of relying on inflation (FT1) and dark
energy (FT2), none of these being physically understood with inflation
and dark energy having no direct evidence by any other (non-cosmological)
experiment. This model thus too does not conserve energy and runs into the
``cosmological energy catastrophe'' (Sec.~\ref{sec:SMoC_tensions}).
Also, it does not address the baryon asymmetry (FT4).

{\it In Summary}, all models that allow the hydrodynamical simulation of
structure formation down to galaxy or galaxy-cluster scales and that
rely on the initial conditions as set by the observed CMB appear to
run into problems. In particular, assuming this initial CMB boundary condition, the SMoC produces too little structure on scales larger than $100\,$Mpc (apart from all the other problems listed in Sec.~\ref{sec:SMoC_tensions}), while the AMoC produces too much of it and structures emerge at too small a redshift.
It appears not possible to perform a SMoC or AMoC
simulation that turns out a model universe which agrees with the
Universe in terms of the large scales and the quick onset of star
formation at $z_{\rm e}>10$ as uncovered by the JWST. Therefore it might seem 
useful to explore models that do
not depend on inflation and on the CMB being the photosphere of a Hot Big
Bang, and to consider suggestions on the origin of the CMB in terms of
other processes.

Such ideas are not new, with Rees (1978 (\cite{Rees1978}, see also \cite{Rowan-Robinson+1979, Wright1982} and references therein) having proposed that the CMB may have been produced by the first star formation with reprocessing of the stellar emission through dust. In the context of the MMoC (the $R_{\rm h}=c\,t$ model), 
Melia (2022, \cite{Melia2022}) suggests the CMB to be the (standard) surface of
last scattering or photosphere produced at $z_{\rm e}\approx 1100$, just as in the SMoC, but the temperature fluctuations in it to stem from $z_{\rm e}\approx 16$ 
as a consequence of the CMB photons scattering on dust produced by the first generations of stars. \cite{Melia2022} notes evidence for this through the 
frequency dependence in the CMB power spectrum observed by the Planck satellite.
Vavrycuk (2018, \cite{Vavrycuk2018}) suggests the CMB to be
entirely dust-reprocessed star light over cosmological distances, given the observed traces of dust between galaxies in the present-day
Universe (i.e., the present-day observed intergalactic dust density leads to an emission over cosmological distances that, according to the calculation presented in \cite{Vavrycuk2018}, is comparable to the observed CMB spectral energy distribution). 
With the possibility that the CMB might have a different origin, it is important to independently assess the validity of a cosmological model.
One such approach would be to obtain a direct measurement of the expansion rate of the Universe (e.g. \cite{Melia2022c} and references therein). 
Another constraint is to obtain an independent measurement of the age of the Universe.

A cosmology-model-independent measurement of the expansion rate of the Universe may be available  using the Cosmic Chronometers method proposed by Jimenez \& Loeb (2002 \cite{JimenezLoeb2002}). The idea is to measure the ages of stellar populations in an ensemble of elliptical (i.e. passively evolving) galaxies at two different redshifts, to obtain a constraint on the time derivative of the redshift, ${\rm d}z_{\rm e}/dt$. The challenge of this method comes with needing to fully understand biases (e.g. Moresco et al. 2020 \cite{Moresco+2020}) that enter through stellar-evolution models, chemical enrichment, the variation of the galaxy-wide IMF, possible on-going star formation at low levels as well as less-massive elliptical galaxies having typically formed later than more massive ones (Kroupa et al. 2013 \cite{Kroupa+2013}, Jerabkova et al. 2018 \cite{Jerabkova+2018}, Yan et al. 2220 \cite{Yan+2021}, Eapen et al. 2022 \cite{Eappen+2022}. A more robust measurement of the actual age of the observed Universe
that is independent of the applied cosmological model can be obtained
from the ages of the oldest known stars.  From table~1 (the entry "{\it Very Metal Poor Stars}") in
Cimatti \& Morsco (2023, \cite{CimattiMoresco2023}), the one-sigma upper limit on the age of the
real Universe is $t_0 = 15.3\,$Gyr (assuming the SMoC: global 
$H_0^{\rm global, stars} = 60.6 \, {\rm km} \, {\rm s}^{-1}\,{\rm Mpc}^{-1}$) while the
three-sigma upper limit is $t_0 = 16.5\,$Gyr (assuming the SMoC: globally
$H_0^{\rm global, stars} = 55.2\, {\rm km} \, {\rm s}^{-1}\,{\rm Mpc}^{-1}$). 
In this context it is interesting to note that Ying et al. (2023 \cite{Ying+23}) measure the absolute age of the globular cluster M92 to be $13.80\pm0.75\,$Gyr in tension with the nominal Planck age ($t_0$, Sec.~\ref{sec:SMoC}) of the universe because the cluster stars are, with [Fe/H]$=-2.30\pm0.10$, not metal-free implying that a previous population of stars must have existed to enrich the interstellar medium from which the cluster formed.

The value of $t_0$ would be larger by a few hundred Myr since star formation can only begin after the birth of the Universe. Such a small value of the $H_0^{\rm global}$ might be 
qualitatively consistent with the larger local value measured using
SN1a as standard candles, in view of the Local Group being in the
$\simgreat 0.6\,$Gpc void (T6, Sec.~\ref{sec:SMoC_tensions}). It will
be necessary to test if such differences between apparent large local
(as measured from the peculiar velocities of galaxies) and small
global expansion rates arise as a consequence of the large matter
inhomogeneities on many Gpc scales as is inherent to the timescape cosmological model, Sec.~\ref{sec:GRbased}).

With the many tensions facing the SMoC and the other models discussed here, the Bohemian model of cosmology
(BMoC) is being explored in Bonn, Prague and Nanjing. The results of
this exploration will be reported in upcoming works, 
but here it is merely
stated that the BMoC avoids inflation, dark matter and dark energy. It rests on 
Milgromian gravitational fields being sourced by matter density
contrasts, just as in the AMoC (Sec.~\ref{sec:AMoC}).  Overall, there are encouraging aspects (expansion rate, 
star-formation histories of galaxies, an older Universe, large density variations over
large scales) and stars can begin to form at a redshift $z_{\rm *form}>20$. 
The first structure formation simulations using the PoR code (Appendix~\ref{sec:MOND}) lead to
positive results in that the BMoC appears to produce a universe full of disk galaxies.

\begin{acknowledgments}
  We thank Eoin {\'O} Colg{\'a}in for useful comments and the staff at
  the Argelander-Institut for Astronomy in Bonn for their kind help
  with the computational system, and Indranil Banik for useful
  discussions. A very small financial support is acknowledged through
  the trans-disciplinary research area TRA-Matter at the University of
  Bonn. We acknowldge the DAAD-Bonn-Prague exchange program at the
  University of Bonn and at Charles University for supporting the
  Bonn-Prague exchange visits.
\end{acknowledgments}

\appendix

\section{Milgromian dynamics (MOND) and the code PoR}
\label{sec:MOND}

Milgromian dynamics has been found to correctly predict (before
the existence of the data) various relations that galaxies are
observed to follow \citep{SandersMcGaugh2002, Scarpa2006, Sanders2007,
  FamaeyMcGaugh2012, Trippe2014, Milgrom2014, Sanders2015,
  Merritt2020, BanikZhao2022}. Additionally, the kinematics of polar-ring galaxies are naturally understood in MOND, while they are not with Newtonian gravitation plus dark matter \cite{Lueghausen+2013}. And, Newtonian gravitation does not account for the properties of the observed tidal tails of open-star-clusters,  while Milgromian gravitation is very successful in doing so \cite{Kroupa+2022}. Finally, Milgromian gravitation appears to have been confirmed also on sub-pc scales using the wide-binary star test \cite{Chae2023}.
  Together with MOND providing equations
of motion, we thus have a strong motivation to investigate
cosmological structure formation in this dynamics theory which is here
taken to be a non-relativistic extension of Newtonian
gravitation.\footnote{Interpretations of MOND as being a
  modified-inertia theory \citep{Milgrom2022} are less well studied
  because they are technically very hard to investigate but are
  reported to be less preferred by the data
  \citep{Chae2022}. Relativistic formulations of MOND have been
  developed and are consistent with gravitational waves travelling at
  the speed of light (\cite{BanikZhao2022} and references
  therein).}

The reader needs to be aware that MOND is an effective framework that encompasses many different explicit formulations of gravitation (e.g. AQUAL vs QUMOND below, see Milgrom 2023 for the latest generalisations \cite{Milgrom2023})). The different explicit formulations lead to subtle differences in the exact acceleration fields. For each such explicit formulation, the interpolating functions also offers degrees of freedom. Hereby it is not meant that  different explicit formulations or interpolating functions can be chosen for different problems at will -- only one such formulation and one interpolating function can be true. Research has not yet allowed the identification of which combination is true though, given the uncertainties in the data and the difficulties in performing calculations in this non-linear theory of dynamics. 

Weak-lensing is often considered to provide evidence for dark matter halos around galaxies and galaxy clusters, but the statistically correlated distortions of background galaxy images by a foreground mass need to be re-evaluated in terms of another gravitational theory, in this case MOND. A bias would emerge due to possible correlations of disk galaxy spins over large regions of spatial volume (e.g. \cite{Shamir2022, ShamirMcAdam2022, KarachentsevZozulia2023, LeeMoon2023}) that would affect the deduced mass distribution using weak lensing analysis if the (erroneous) assumption is made that galaxies are uncorrelated.  The correct distribution of mass as deduced by weak-lensing surveys will then be different to that arrived at by using the (invalid) SMoC. The expected distribution of lensing mass as calculated in MOND of the galaxies in the nearby Universe has been documented by Oria et al. (2021 \cite{Oria+2021}), who point out that regions with effective negative Newtonian mass density are expected to exist in this gravitational framework. 

\subsection{MOND}

With the baryonic mass density at the position vector, $\vec{R}$,
being $\rho_{\rm b}(\vec{R})$, the standard (Newtonian) Poisson
equation reads
\begin{equation}
  \vec{\nabla} \cdot \left[ \vec{\nabla} \Phi_{\rm N} \right] = 4\pi \, G \,  \rho_{\rm b},
\label{eq:Poisson}
\end{equation}
where $\Phi_{\rm N}(\vec{R})$ is the Newtonian potential, the negative
gradient of which provides the acceleration at $\vec{R}$.  A
generalised dynamics can be considered by rewriting the left hand side
in terms of the well known p-Laplace operator,
\begin{equation}
  \vec{\nabla} \cdot \left[  \left( \frac{ |\vec{\nabla} \Phi_p|
        }{ a_o}\right)^{p-2}  \,  \vec{\nabla} \Phi_p
  \right] = 4\pi \, G \, \rho_{\rm b},
\label{eq:2p}
\end{equation}
the constant, $a_0$, with units of acceleration being necessary for
dimensional reasons.  This is a quasilinear elliptic partial
differential equation of second order.
In the case $p=di$, where $di$ is the dimension of the domain, the corresponding
$p-$Dirichlet integral is conformally invariant (e.g. \cite{Lindqvist2019}). Interestingly, this is related to the space-time scale invariance of the SIV model and of MOND \cite{Milgrom2009, MaederGueorguiev2020}. 
For $p=2$ the standard Poisson equation (Eq.~\ref{eq:Poisson}) is obtained that underlies
Newtonian gravitation ($\Phi_{\rm N} = \Phi_p$). 
The particular case $p=3$ yields a non-linear gravitational theory which corresponds
to the deep-MOND limit, when $|\vec{\nabla} \Phi_{p=3}| \ll a_0$
where $\Phi_{p=3}=\Phi_{\rm M}(\vec{R})$ is the Milgromian
potential, the negative gradient of which provides the acceleration at
$\vec{R}$.

The transition from the $p=3$
case to the $p=2$ ($|\vec{\nabla} \Phi_{\rm p}| \gg a_0$) case can be
described by inserting an interpolating function,
$\mu(x) \rightarrow 1$ for
$x= |\vec{\nabla}\Phi| / a_0 \rightarrow \infty$ and
$\mu(x) \rightarrow x$ for $x \rightarrow 0$. The full
Milgromian-Poisson equation thus becomes
\begin{equation}
\vec{\nabla} \cdot 
\left[
\mu\left( \frac{|\vec{\nabla}\Phi_{\rm M}| }{ a_0} \right) \,
\vec{\nabla}\Phi_{\rm M}
\right] = 4\,\pi\,G\,\rho_{\rm b}.
\label{eq:genPois1}
\end{equation}
Eq.~\ref{eq:genPois1} can be derived from a quadratic Lagrangian
(AQUAL) as a generalisation of the non-relativistic Newtonian
Lagrangian \citep{BekensteinMilgrom1984}.  Thus we have energy- and
momentum conserving time-integrable equations of motion of
non-relativistic gravitating bodies in Milgromian dynamics.

By the properties of the divergence operator, the left hand side of
Eq.~\ref{eq:genPois1} can be written
\begin{equation}
\vec{\nabla} \cdot 
\left[
\mu\left( \frac{|\vec{\nabla}\Phi_{\rm M}| }{ a_0} \right) \,
\vec{\nabla}\Phi_{\rm M}
\right] = \vec{\nabla}\mu(x) \cdot \vec{\nabla}\Phi_{\rm M} + \mu(x)
\;   \left(\vec{\nabla} \cdot  \vec{\nabla} \Phi_{\rm M} \right),
\label{eq:LHS}
\end{equation}
such that
\begin{equation}
 \vec{\nabla} \cdot  \left[ \vec{\nabla} \Phi_{\rm M} \right]
 = 4\,\pi\,G \; \left[ \frac{1 }{ \mu(x)}     \left(
      \rho_{\rm b}
-\frac{ \vec{\nabla}\mu(x) \cdot \vec{\nabla}\Phi_{\rm M} }{ 4\,\pi\,G }
\right) \right] \, ,
\label{eq:phantomDM1}
\end{equation}
where the square brackets can be interpreted to be the total mass
sourcing the Milgromian potential, $\Phi_{\rm M}$. This total mass is
composed of the baryonic matter, $\rho_{\rm b}$, modified by $\mu$ and
the remaining term can be referred to as phantom dark matter (PDM)
which however is not made up of real particles and does not lead to
dynamical dissipation or friction. This formulation of Milgromian
gravitation cannot be solved easily for $\Phi_{\rm M}$ given
$\rho_{\rm b}$, because $\Phi_{\rm M}$ is found in the dynamics term
(the left-hand side) and also on the 
right-hand side (the source term) of Eq.~\ref{eq:phantomDM1}.

This difficulty and the similarity of Eq.~\ref{eq:phantomDM1} 
with the standard Poisson equation (Eq.~\ref{eq:Poisson}) leads 
to the motivation to develop a
more direct approach.  Essentially, this quasi-linear MOND (QUMOND) 
approach replaces the right-hand side of Eq.~\ref{eq:phantomDM1} by 
an approximation based on the Newtonian potential. 
A new Lagrangian leads to the QUMOND formulation \citep{Milgrom2010} 
that allows the calculation
of $\Phi_{\rm M}$ more easily. According to QMOND,
\begin{equation}
  \vec{\nabla} \cdot \vec{\nabla} \Phi_{\rm QM} (\vec{R})
  =
  4 \pi G \; \left( \rho_{\rm b}(\vec{R}) + \rho_{\rm ph}(\vec{R}) \right) \, ,
\label{eq:genPois2}
\end{equation}
where $\rho_{\rm ph}(\vec{R})$ is the phantom dark matter
density (PDM). Therefore the total Milgromian gravitational potential,
$\Phi_{\rm QM} = \Phi_{\rm N} + \Phi_{\rm ph}$, can be split into a
Newtonian part, $\Phi_{\rm N}$, and an additional phantom part,
$\Phi_{\rm ph}$.  The matter density distribution,
$\rho_{\rm ph}(\vec{R})$, that would, in Newtonian gravitation, yield
the additional potential, $\Phi_{\rm ph}(\vec{R})$, and therefore
obeys
$\nabla^2 \Phi_{\rm ph}(\vec{R}) = 4\pi G \rho_{\rm ph}(\vec{R})$, is
known as the PDM density,
\begin{equation}
	\rho_{\rm ph}(\vec{R}) = \frac{1} {4\pi G}  \vec{\nabla} \cdot 
\left[   \widetilde{\nu}\left( \frac{|\vec{\nabla} \Phi_{\rm
      N}(\vec{R})| }{ a_0} \right)\vec{\nabla} \Phi_{\rm N}(\vec{R}) 
\right]  \,,
\label{eq:phantomDM2}
\end{equation}
where $\widetilde{\nu}(y) \rightarrow 0$ for $y \gg 1$ (Newtonian
regime) and $\widetilde{\nu}(y) \rightarrow y^{-1/2}$ for $y \ll 1$
(Milgromian regime), Note that $\mu$ and $\widetilde{\nu}$ are
algebraically related (see \cite{FamaeyMcGaugh2012, Milgrom2014})
with $y=|{\vec{\nabla}} \Phi_{\rm N}|/a_0$. Remember that PDM is
merely a mathematical aid for calculating the additional gravity in
Milgrom's formulation, giving it an analogy in Newtonian dynamics. The
PDM can become negative locally \citep{Candlish2016}. It is important
to keep in mind that $\Phi_{\rm M}$ and $\Phi_{\rm QM}$ are not the
same and subtle differences may emerge in the transition regime that
will need scrutiny in the future.

The PDM density that would source the Q-Milgromian force field in
Newtonian gravity can thus be calculated straightforwardly using
standard grid-based methods (see eq.~35 in
\cite{FamaeyMcGaugh2012}, and also \cite{Gentile+2011};
\cite{Milgrom2010}; \cite{Lueghausen+2013};
\cite{Lueghausen+2014}, \cite{Lueghausen+2015}).  Starting with
the known baryonic density distribution, $\rho_{\rm b}(\vec{R})$,
$\Phi_{\rm N}$ is obtained by solving the standard Poisson equation
(Eq.~\ref{eq:Poisson}).  Thereafter and at each time step, the new PDM
distribution is calculated via Eq.~\ref{eq:phantomDM2}, which, in sum
with $\rho_{\rm b}$, then yields the full Milgromian potential
(Eq.~\ref{eq:genPois2}).

An important consequence of the non-linear nature of Milgromian
gravitation is that an external gravitational field (e.g. from a
galaxy cluster) reduces $\rho_{\rm ph}$ of the system (e.g. a galaxy)
immersed in this external field (e.g. \cite{BradaMilgrom2000,
  WuKroupa2013, Candlish+2018}).  The system looses most (but not all)
of its non-Newtonian gravitational mass and becomes more Newtonian
(the ``external field effect'' or EFE, see the reviews on MOND above
and \cite{Kroupa+2022} for a discussion). The implication of the
EFE for cosmological structure formation is important: overdensities
inhibit structure growth in their vicinity therewith enhancing density
differences \citep{Haslbauer+2020}.

\subsection{The PoR code}

The above QUMOND technique was implemented independently by Lueghausen et al. (2015 
\cite{Lueghausen+2015}) and Candlish et al. (2015 \cite{Candlish+2015}) into the existing {\sc
  RAMSES} code developed for Newtonian gravitation by
Teyssier (2002 \cite{Teyssier2002}). {\sc RAMSES} employs an adaptively refined grid
structure in Cartesian coordinates such that regions of higher density
are automatically resolved with a higher resolution.  In the {\sc Phantom
of Ramses} (PoR) code, \cite{Lueghausen+2015} added a subroutine which,
on the adaptive grid, computes the PDM density from the Newtonian
potential (Eq.~\ref{eq:phantomDM2}) and adds this (mathematical)
DM-equivalent density to the baryonic one in terms of particles. These
PDM particles are replaced by new PDM particles in the next time step,
i.e., they do not carry an ID and are not integrated forward in time.
The Poisson equation is then solved again to obtain the Milgromian
potential, $\Phi_{\rm QM} = \Phi_{\rm N}+\Phi_{\rm ph}$, i.e., the
``true'' (Milgromian) potential, which is used to integrate the
stellar particles in time through space (each stellar particle
experiencing the acceleration
$\vec{a}_{\rm star}(\vec{R}) = -\vec{\nabla}\Phi_{\rm M}(\vec{R})$)
and to solve the Euler equations for the dynamics of the gas.

It is to be noted that the AQUAL and QUMOND formulations are different
theories complying to the MOND paradigm \citep{Milgrom2014}. Thus,
simulations with the AQUAL version (e.g. using the
Candlish et al. 2015 \cite{Candlish+2015} code) will show subtle differences to
simulations relying on the QUMOND formulation by Lueghausen et al. (2015 \cite{Lueghausen+2015}). 
Which formulation of
MOND is closer to reality will need to be studied in the future (see
Chae et al. (2022 \cite{Chae+2022}). Here we concentrate on the QUMOND version coded
into PoR as the first step towards a very novel computable
cosmological model.

The PoR code adopts as the default interpolating function
\begin{equation}
  \widetilde{\nu}(y) = -\frac{1 }{ 2} + \left( \frac{1}{ 4} + \frac{1 }{ y}
  \right)^\frac{1}{ 2} \, ,
\label{eq:interpol}
\end{equation}
which has already been used on a number of problems (for the manual and
description see Nagesh et al. (2021 \cite{Nagesh+2021}). Other interpolating functions
can be employed subject to observationally-given constraints
(e.g. \cite{Hees+2016}).

The PoR code includes sub-grid baryonic physics algorithms that allow
different degrees of realism in the description of star formation.
Simulations have been done to model Antennae-like interacting galaxies
\cite{Renaud+2016}, the formation of exponential disk galaxies
\cite{Wittenburg+2020}, the formation of elliptical galaxies
\cite{Eappen+2022} as well as of $10^7\,M_\odot$ to
$10^{11}\,M_\odot$ heavy galaxies on the main sequence
\cite{Nagesh+2023}. These detailed tests of how the properties of a
galaxy change with increased complexity of the in-build sub-grid
baryonic processes have shown that galaxies evolved in MOND are not
sensitive to these. In this work star-formation is turned off because
the resolution achieved does not allow the formation of stellar
particles to have a physical meaning. The PoR code has also been used
without enabling star-formation to model the Milky-Way--Andromeda
encounter which produced the disks of satellite galaxies around both
\cite{Bilek+2018, Banik+2022} and also to study the global stability of
the M33 galaxy \citep{Banik+2020}.

\subsection{Cosmological simulations with PoR}

Nusser (2002 \cite{Nusser2002}) details the implementation of the MOND field
equation (Eq.~\ref{eq:genPois1} and~\ref{eq:genPois2}) into
computations of structure formation in an expanding universe.  QUMOND
has been applied in cosmological structure formation simulations using
particles by Katz et al. (2013 \cite{Katz+2013}) and Angus et al. (2013 \cite{Angus+2013}) 
employing a
code developed by Llinares et al. (2008 \cite{Llinares+2008}) that accounts for co-moving
coordinates in an expanding universe.

When including gas dynamical processes in order to compute how
structures form and evolve in an expanding space, the equations of
motion are here integrated in super-co-moving coordinates tailored
specifically for an ideal gas with the ratio of specific heats
being~5/3 \cite{MartelShapiro1998}.  This is of advantage because, as
\cite{MartelShapiro1998} argue, in the absence of structure and in an
expanding box, in co-moving coordinates the density is constant in
time and the mass elements are at rest. In super-co-moving variables
this is also true, and in addition, a gas with a ratio of specific
heats as above also has its thermodynamic variables remain constant
when structure is absent.  RAMSES \citep{Teyssier2002}, and thus also
PoR, adopts super-co-moving coordinates. The details of how QUMOND is
implemented on calculating the Milgromian potential, $\Phi_{\rm QM}$,
from the density difference relative to the mean density, are
available in Candlish (2016 \cite{Candlish2016}) and Wittenburg et al. (2023 \cite{Wittenburg+2023}).

\bibliographystyle{JHEP}
\bibliography{refs_PKcorfu}{}
%

\end{document}